\pdfoutput=1
\documentclass[acmsmall,screen]{acmart}

\usepackage{fancyhdr}

\usepackage[capitalize]{cleveref}
\usepackage{upgreek}
\usepackage[frozencache]{minted}
\setminted{fontsize=\small}

\AtBeginDocument{%
  }

\setcopyright{none}
\copyrightyear{2026}
\acmYear{2026}
\acmDOI{}

\newcommand{\idx}[1]{\,{\scriptsize(\##1)}}
\newcommand{\miniKanren}{\textsc{miniKanren}}
\newcommand{\muKanren}{\textsc{$\upmu$Kanren}}
\newcommand{\conde}{\textbf{cond}$^e$}

\begin{document}

\title{Towards Bottom-Up Enumeration in \miniKanren{} via Pruning and Memoization}

\author{Nikolai Kudasov}
\affiliation{%
  \institution{Innopolis University}
  \country{Russia}}
\email{n.kudasov@innopolis.ru}

\renewcommand{\shortauthors}{N.~Kudasov}

\settopmatter{printacmref=false}
\settopmatter{printfolios=true}
\renewcommand\footnotetextcopyrightpermission[1]{}
\pagestyle{fancy}
\fancyfoot{}
\fancyfoot[R]{miniKanren'26}
\fancypagestyle{firstfancy}{
  \fancyhead{}
  \fancyhead[R]{miniKanren'26}
  \fancyfoot{}
}
\makeatletter
\let\@authorsaddresses\@empty
\makeatother

\begin{abstract}
  We present two small library combinators on top of plain \miniKanren{},
  designed to bring bottom-up enumeration with observational
  deduplication, the standard tool in non-relational
  program-by-example (PBE) synthesizers, into the relational setting.
  The first combinator, \texttt{prune}, deduplicates an answer stream
  by a user-supplied key, typically the input/output behavior of the
  candidate. The second, \texttt{defrel/bank}, memoizes a relation
  against canonical fresh variables so that a single pruned answer
  stream is built bottom-up and replayed at every call site.
  We also discuss a weighted variant, \texttt{defrel/bank-w}, which
  attaches admissible upper bounds to immature streams to recover
  best-first enumeration in cases where the natural depth-first
  canonical order misses compact representatives. On a preliminary
  PBE benchmark of arithmetic and string synthesis targets,
  \texttt{defrel/bank} substantially outperforms the depth-bounded
  baseline on most deep targets, while losing on a small family
  where the canonical depth-first enumeration order misses compact
  representatives. We leave a broader empirical evaluation to an
  extended version of this paper.
\end{abstract}


\maketitle
\thispagestyle{firstfancy}

\section{Introduction}
\label{sec:intro}

Bottom-up enumeration with observational deduplication is the
standard approach in modern program-by-example (PBE)
synthesizers~\cite{albarghouthi-2013,eusolver-2017,probe-2020,bustle-2021}.
The recipe is to maintain a growing worklist of candidate
expressions, deduplicate the worklist by behavior on the example
inputs, and extend it level by level until some candidate matches
the specification. While this approach has been very effective in
imperative and functional settings,
it sits awkwardly in \miniKanren{}-style relational
programming~\cite{byrd-thesis-2009,hemann-friedman-2013}, where the
natural search discipline is top-down resolution against a fair
stream of candidates.

Indeed, several \miniKanren{} recipes have been developed for
PBE-style synthesis. The dominant one is to write a relational
interpreter for the target language and run it ``backwards'' against
the input/output examples~\cite{byrd-ballantyne-rosenblatt-might-2017,barliman};
subsequent work has pursued, among other directions, neural
guidance~\cite{neuralkanren-2018}, multi-stage
programming~\cite{ballantyne-multistage-2025}, and concrete
applications such as pattern-matching
compilation~\cite{kosarev-lozov-boulytchev-2020},
\textsc{JavaScript} synthesis~\cite{chirkov-rosenblatt-might-zhang-2020},
and type
inference~\cite{domoratskiy-boulytchev-2024}. A simpler
enumerate-and-test recipe writes a relation \texttt{(expr e)} that
enumerates candidate expressions via \conde{}, conjoins it with a
predicate that tests \texttt{e} against the input/output examples by
unification, and issues a \texttt{run~1} query to obtain the first
matching candidate. Both recipes work at toy depths (or narrow search spaces) and quickly
drown at realistic ones. The underlying issue, in the
enumerate-and-test recipe that we focus on in this paper, is that
the search tree contains many behaviorally-equivalent expressions,
and a fresh deduplication table created at each recursive level
cannot reuse the work performed at the previous level.

We aim to close this gap with two small library combinators on top of
an unmodified \muKanren{} core. The first, \texttt{prune}, filters an
answer stream by a user-supplied key (typically the input/output
behavior of a candidate), keeping one representative per equivalence
class. The second, \texttt{defrel/bank}, memoizes a relation
against canonical fresh variables, so that a single pruned answer
stream is built \emph{once} per \texttt{(run \ldots)} and replayed
at every call site. Together, these combinators recover bottom-up enumeration
without leaving the relational setting and without changing the host
search discipline. \Cref{fig:teaser} shows the resulting API: a
complete arithmetic-PBE expression enumerator in six lines, with no
depth parameter and no per-call deduplication plumbing.

\begin{figure}[t]
\begin{minted}{racket}
(defrel/bank (arith-bank e)
  #:prune (arith-key e)
  (conde
    [(== e 'x)] [(== e 0)] [(== e 1)]
    [(fresh (l r)
       (conde [(== e `(plus ,l ,r))]
              [(== e `(times ,l ,r))])
       (arith-bank l) (arith-bank r))]))
\end{minted}
\caption{An arithmetic-PBE expression enumerator using
\texttt{defrel/bank}. The \texttt{\#:prune} clause attaches an
observational key to the relation; the body is ordinary \miniKanren{}
with no depth parameter. The depth-bounded equivalent (in~\Cref{fig:arith-bounded}) requires
an extra integer argument, a base case, and an explicit
\texttt{(prune key \ldots)} wrapper at every recursive level.}
\label{fig:teaser}
\end{figure}

\subsection{Contribution}

Specifically, our contribution is as follows:
\begin{enumerate}
  \item In \cref{sec:prune}, we introduce \texttt{prune}, an
    answer-stream deduplication combinator keyed on a user-supplied
    function, together with the \texttt{ground-key} and
    \texttt{when-ground} helpers that lift term-level functions on
    ground terms into prune keys.
  \item In \cref{sec:bank}, we present \texttt{defrel/memo} and
    \texttt{defrel/bank}, two combinators for relation memoization
    against canonical fresh variables. The latter further provides
    bottom-up enumeration with shared deduplication across all call
    sites of a relation. We also discuss the lazy-template and
    thunk-collapse tricks that allow recursive memoized relations to
    perform competitively.
  \item In \cref{sec:bankw}, we discuss \texttt{defrel/bank-w}, a
    depth-decayed best-first variant whose weights live on
    \emph{immature} streams. This variant uses the lazy-thunk weight
    ceiling as an A*-style admissible heuristic, allowing us to avoid
    forcing the very recursion the memoized stream is in the middle of
    building.
  \item In \cref{sec:eval}, we report a preliminary empirical
    evaluation of all three variants on arithmetic and string PBE
    benchmarks up to depth~6. In particular, \texttt{defrel/bank} is
    9--99$\times$ faster than depth-bounded \miniKanren{} on 6 out
    of 8 deep arithmetic targets, and we characterize the two cases
    it loses. We keep the benchmark suite small
    and plan a more thorough evaluation,
    including additional PBE domains and a comparison with bottom-up
    SyGuS\footnote{Syntax-Guided Synthesis} solvers, for an extended version of this paper.
\end{enumerate}

The implementation of our combinators, together with the benchmark
drivers used to produce \cref{tab:eval}, is included in the Appendix
(\cref{sec:appendix-supp}) and maintained at
\url{https://github.com/fizruk/prune-kanren}.

\section{Background}
\label{sec:background}

In this section, we briefly recall the standard \muKanren{} core
that we build upon, set up the program-by-example setting we use as
a running example throughout the paper, and explain why a na\"ive
\miniKanren{} enumeration blows up at realistic depths.

\subsection{The \muKanren{} Core}
\label{sec:background-mk}

We assume the standard \muKanren{} core of Hemann and
Friedman~\cite{hemann-friedman-2013}: equality \texttt{==},
fresh-variable introduction \texttt{call/fresh}, disjunction
\texttt{disj}, conjunction \texttt{conj}, and an inverse-$\eta$
delay\footnote{Sometimes referred to as \texttt{Zzz} in the
literature.} that schedules recursive calls fairly. Recall that, in
this setting, a \emph{goal} is a function from a state (a
substitution paired with a fresh-variable counter) to a stream of
states, that is, of type \texttt{state $\to$ stream of states}. Each
state in the resulting stream represents one way in which the goal
succeeds, recording the substitution extensions and any new fresh
variables introduced along the way. Answer streams themselves are
either finite lists, immature thunks, or interleavings thereof, and
the operator \texttt{pull} repeatedly forces thunks until a cons
cell or an empty cell is exposed. Variadic surface forms such as
\texttt{conde}, \texttt{fresh}, \texttt{run}, and \texttt{run*} are
defined as wrappers over these primitives in the usual way.

It is important to note that we work with this minimal core
deliberately: our prototype supports neither disequality
constraints, nor symbolic or finite-domain constraints, nor any
other extension of the standard \muKanren{} substitution. Equality
constraints, introduced by \texttt{==} and resolved by ordinary
unification, are the only constraints that goals in our setting can
impose on a state. The combinators of \cref{sec:prune,sec:bank,sec:bankw}
are designed against this minimal core.

We believe this restriction is a matter of engineering rather than
of principle. On the memoization side, replay (\cref{sec:bank})
renames canonical bindings into the caller's namespace, and the same
renaming could be applied to a constraint store attached to each
canonical cell. On the pruning side, keys computed from ground
terms that capture the entire answer, as in our PBE examples, appear
unproblematic. In general, however, more care is needed. Two states
that agree on the prune key may carry different constraint stores.
Thus, dropping one of them may lose answers unless the key is chosen
to respect the constraints. We have not implemented either part, and
leave the interaction of pruning and memoization with richer
constraint stores to future work.

Our reimplementation of \muKanren{} is in the modules \texttt{microkanren.rkt}
and \texttt{wrappers.rkt}, listed in
\cref{sec:appendix-microkanren,sec:appendix-wrappers}.

\subsection{Programming by Example}
\label{sec:background-pbe}

Programming by example (PBE) is a form of program synthesis in which
the specification is a small set of input/output examples rather
than a full functional specification~\cite{gulwani-2011}. The
synthesizer fixes a grammar of candidate programs and searches for
\emph{any} program that agrees with every example. For instance,
given the examples $2 \mapsto 4$, $3 \mapsto 9$, and $4 \mapsto 16$
over the arithmetic grammar of \cref{fig:teaser}, a PBE synthesizer
is expected to return \texttt{(times x x)}. Any expression with the
same behavior, such as \texttt{(times x (times x 1))}, is equally
acceptable, since the examples do not distinguish them. Indeed, a
handful of examples underdetermines the target. Thus, PBE systems
typically return the first program found, and the user refines the
examples if the result is not the intended one. The flagship
application of this workflow is string transformation in
spreadsheets~\cite{gulwani-2011}, where an end user supplies a
couple of before/after pairs instead of writing a formula. The
string domain in \cref{sec:eval} is in this style.

Throughout the paper, we use a single running setting: synthesize an
arithmetic expression over a single variable $x$ and the constants 0
and 1 from a handful of input/output examples. The grammar of such
expressions is given in \cref{fig:teaser}, with atoms \texttt{x},
\texttt{0}, and \texttt{1}, and the binary operators \texttt{plus}
and \texttt{times}. A relation \texttt{(expr e)} enumerates candidate
expression terms; a predicate \texttt{(matches e \textit{io})}
succeeds when the term \texttt{e}, evaluated on each example input
in \textit{io}, produces the expected output. The expression

\begin{center}\small
  \texttt{(run 1 (e) (expr e) (matches e \textit{io}))}
\end{center}

\noindent then returns the first matching candidate. Our target
set, drawn from the benchmark modules
(\cref{sec:appendix-shallow-bench,sec:appendix-deep-bench}), ranges
from $x^2$ (with examples $(2,4)$, $(3,9)$, $(4,16)$) up through
$x^7$ and $(1+x)^5$.

In many uses of
relational programming, e.g. when running a relational interpreter
backwards, one is interested in the full multiplicity of answers,
and collapsing behaviorally equivalent programs would be
inappropriate. The PBE query above is different. It asks for
\emph{some} program consistent with the examples. Thus, one
representative per behavior suffices, and the remaining members of
each equivalence class are pure search overhead. For these reasons, our combinators are strictly opt-in.
Deduplication applies only to relations that the user explicitly
wraps in \texttt{prune} or defines via \texttt{defrel/bank}, and
only with respect to the equivalence the user supplies. The rest of
the program retains the usual \miniKanren{} semantics.
\Cref{sec:prune} makes the induced completeness guarantee precise.

\subsection{Why Na\"ive Enumeration Blows Up}
\label{sec:background-blowup}

The number of syntactically distinct expression terms grows
exponentially in the depth bound, but the number of distinct
\emph{behaviors}, that is, output tuples on the example inputs,
grows much more slowly. By depth~3, over the grammar of
\cref{fig:teaser}, there are several thousand candidate terms but
only a few dozen distinct behaviors. For example, the terms
\texttt{(times x x)}, \texttt{(times (times x 1) x)}, \texttt{(times
x (times x 1))}, \texttt{(times 1 (times x x))}, and many others all
compute the same outputs on any input, and are therefore
interchangeable for any PBE query. A search that prunes by syntactic
equality misses these collapses entirely. What we need is
\emph{observational} deduplication, keyed by the behavior tuple
rather than by the syntactic shape of the term. This is the main
trick of bottom-up
synthesizers~\cite{albarghouthi-2013,probe-2020,bustle-2021,eusolver-2017}.
In the next two sections, we transplant it into \miniKanren{}.

\section{The \texttt{prune} Combinator}
\label{sec:prune}

The first combinator we introduce, \texttt{prune}, wraps a goal and
filters its answer stream so that at most one state is emitted per
distinct value of a user-supplied \emph{key} function. Specifically,
its signature is

\begin{center}
  \texttt{prune : (state $\to$ any) $\to$ goal $\to$ goal}.
\end{center}

\noindent The equivalence used to prune is chosen \emph{per call}, so
that the user is free to pick the equivalence appropriate to the
problem at hand. Typical choices include the behavior on the example
inputs for PBE, the shape modulo $\alpha$-renaming for relational
interpreters, and syntactic identity for plain deduplication.
\Cref{fig:prune-impl} gives the entire implementation of
\texttt{prune}.

\begin{figure}[t]
\begin{minted}{racket}
(define skip-prune 'skip-prune)

(define (prune key g)
  (lambda (s/c)
    (prune-stream key (make-hash) (g s/c))))

(define (prune-stream key seen $)
  (cond
    [(null? $) '()]
    [(procedure? $) (lambda () (prune-stream key seen ($)))]
    [else
     (let* ([s/c (car $)] [k (key s/c)])
       (cond
         [(eq? k skip-prune)
          (cons s/c (prune-stream key seen (cdr $)))]
         [(hash-has-key? seen k)
          (prune-stream key seen (cdr $))]
         [else
          (hash-set! seen k #t)
          (cons s/c (prune-stream key seen (cdr $)))]))]))
\end{minted}
\caption{The \texttt{prune} combinator. The deduplication table is
created fresh on each call and captured by the stream's lazy thunks,
so the deduplication state survives the inverse-$\eta$ delay used by
\texttt{mplus} and \texttt{bind}.}
\label{fig:prune-impl}
\end{figure}

\subsection{The \texttt{skip-prune} Sentinel}

A key function that depends on the value of a logic variable cannot
produce a meaningful key when that variable is still fresh.
To handle this case, we introduce a sentinel value \texttt{skip-prune},
which tells \texttt{prune-stream} to emit the current state without
recording it; that state will be revisited later, once the variable
is bound. Two small helpers package the common case in which the key
is computed from a term once it has been instantiated.
First, \texttt{ground-key v f} walks \texttt{v} in the current
substitution and applies \texttt{f} if the result is ground, returning
\texttt{skip-prune} otherwise. Second, \texttt{when-ground v pred} is
the predicate-as-goal companion that succeeds when \texttt{v} walks
to a ground term satisfying \texttt{pred}.

Note that a state is subject to deduplication only when its key is
computed, and \texttt{ground-key} computes a key only when the term
is ground. Thus, candidate terms that still contain logic variables
are never recorded in the table and never dropped. They pass through
\texttt{prune} unfiltered and become subject to deduplication only
once instantiated. In particular, while the enumerators in this
paper happen to produce fully ground candidates, \texttt{prune}
itself does not rely on this property.

\subsection{A Worked Example}

For the arithmetic-PBE setting of \cref{sec:background-pbe}, the
behavior key takes the form

\begin{center}\small
  \texttt{(ground-key e (lambda (t) (map (lambda (i) (arith-interp t i)) inputs)))},
\end{center}

\noindent where \texttt{inputs} is the list of example inputs (for
instance, $(2, 3, 4)$), and \texttt{arith-interp} is the host (non-relational)
interpreter for arithmetic expressions. The depth-bounded enumerator
that we use as a baseline throughout the paper is shown in
\cref{fig:arith-bounded}. Wrapping it with the behavior key above
keeps exactly one representative per behavior tuple: of the many
depth-3 terms that compute $(4, 9, 16)$ for $x \in \{2, 3, 4\}$,
only the first one encountered is emitted, and the rest are silently
dropped.

\begin{figure}[t]
\begin{minted}{racket}
(define (arith-bounded e depth)
  (prune (arith-key e)
    (cond
      [(zero? depth)
       (conde [(== e 'x)] [(== e 0)] [(== e 1)])]
      [else
       (conde
         [(== e 'x)] [(== e 0)] [(== e 1)]
         [(fresh (l r)
            (conde [(== e `(plus ,l ,r))]
                   [(== e `(times ,l ,r))])
            (arith-bounded l (- depth 1))
            (arith-bounded r (- depth 1)))])])))
\end{minted}
\caption{The depth-bounded baseline enumerator for arithmetic PBE.
The integer \texttt{depth} is decremented at each recursive call and
a base case is taken when it reaches zero. The body is wrapped in
\texttt{(prune (arith-key e) \ldots)} at every level, so a fresh
deduplication hash is allocated at every recursive call site. We use
this enumerator as the \texttt{bounded} engine in our
evaluation (\cref{sec:eval}).}
\label{fig:arith-bounded}
\end{figure}

\subsection{Completeness}

Pruning never removes an equivalence class, only its non-first
representatives. That is, for each key value, \texttt{prune-stream}
emits the first state and drops the subsequent states with the same
key. Therefore, for every state \texttt{s} of the unpruned
goal there is a state \texttt{s$'$} of the pruned goal with
\texttt{key(s$'$) = key(s)}. A caller that observes states only
through \texttt{key} (which is the case for PBE, where
\texttt{matches} consumes only the behavior tuple) thus sees a
complete answer set up to the equivalence induced by \texttt{key}.

\subsection{Why the Hash is Per-Call}
\label{sec:prune-shared}

Using \texttt{prune} in a recursive definition (e.g. in \Cref{fig:arith-bounded})
leads to many calls to \texttt{prune}, each with its own hash table
and deduplication work.
A natural-looking optimization is to thread a single deduplication
table across all nested \texttt{prune}s via a parameter, in the hope
of amortizing the work across recursive levels. Unfortunately, this
breaks completeness. With one shared table, the outer prune emitting
\texttt{(== e 'x)} marks the behavior $(2, 3, 4)$ as seen; the
recursive subgoal \texttt{(arith-bounded l \ldots)} can then no longer emit
\texttt{l = 'x}, because that same behavior is already in the table.
As a consequence, the candidate \texttt{(times x x)}, which depends
on \texttt{l = 'x}, becomes unreachable. In other words, the shared
hash captures ``seen as the outer \texttt{e}'' and conflates it
with ``seen as the inner \texttt{l}''. A per-call hash avoids this
conflation. We preserve a witness of this issue in the module
\texttt{shared-table-witness.rkt} listed in
\cref{sec:appendix-witness-listing}. The
relational generalization that we present in \cref{sec:bank}
sidesteps the issue entirely by pruning at the level of
\emph{canonical} variables, where the variable identity is fixed.

\section{From \texttt{prune} to Bottom-Up Enumeration: \texttt{defrel/bank}}
\label{sec:bank}

Wrapping a depth-bounded enumerator with \texttt{prune}
(\cref{sec:prune}) deduplicates within each call, but the
deduplication work is not shared across calls. In the arithmetic
recursion of \cref{fig:arith-bounded}, the outer prune over \texttt{(arith-bounded e
depth)} and the recursive prune over \texttt{(arith-bounded l (- depth 1))} each
allocate a fresh hash, and the depth-1 subgoal re-pruning the same
expression terms that the depth-2 subgoal just pruned is, in fact,
pure overhead. What we want instead is a single stream of
representatives (a ``bank''), built once, shared across all call
sites of the relation, and grown bottom-up as new representatives
become available. This is the standard layout of non-relational
bottom-up
synthesizers~\cite{albarghouthi-2013,eusolver-2017,probe-2020,bustle-2021},
and we obtain it in \miniKanren{} by memoizing the relation against
\emph{canonical} fresh variables.

\subsection{Canonical Variables and Replay with Renaming}

Our first idea is to memoize a relation against \emph{canonical
variables}. For a relation of arity $N$, the canonical variables are
its parameters, instantiated as the first $N$ fresh variables
\texttt{(var 0)}, \ldots, \texttt{(var N-1)} of an otherwise empty
state. Intuitively, they represent the \emph{most general call} of
the relation. Running the body once against the canonical variables
produces the \emph{canonical stream}, that is, the answer stream of
the most general call. Its cells are state/counter pairs in which
the parameters are bound to answer terms, with internal fresh
variables counting upward from $N$. We refer to one such cell as a
\emph{canonical cell}. Readers familiar with tabled
\miniKanren{}~\cite[Part~IV]{byrd-thesis-2009}
may read a canonical cell as the analogue of a cache entry, for a
cache keyed on the most general call. We return to this analogy in
\cref{sec:related}. The stream is stored in a per-\texttt{(run \ldots)} cache,
which we expose as the \texttt{current-memo} parameter; each
subsequent call \texttt{(rel $a_1$ \ldots $a_N$)} then replays the
canonical stream against the caller's arguments. The replay is
itself a stream of caller states, computed cell by cell. For each
canonical cell $(\sigma_c, c_c)$ and each input position $i \in
\{0, \ldots, N-1\}$, the canonical value $\sigma_c$ assigns to
\texttt{(var i)} is walked, then renamed (input variables
\texttt{(var i)} for $i < N$ are replaced by the caller's
$a_i$, and canonical internal fresh variables \texttt{(var j)} for
$j \geq N$ are shifted into the caller's namespace by adding the
caller's counter offset), and finally unified into the caller's
substitution. A canonical cell survives the replay only when every
renamed value unifies with its corresponding caller argument;
otherwise that cell is silently dropped. As a consequence,
structured caller arguments such as \texttt{(times a b)} are handled
by unification ``for free'': if the canonical value at position $i$
walks to \texttt{(times $l_c$ $r_c$)} for some $l_c, r_c$, then
unifying against \texttt{(times a b)} either succeeds and extends
the caller's substitution with bindings for \texttt{a} and
\texttt{b}, or fails because the heads disagree. No special case is
needed.

To make this concrete, consider running the relation \texttt{arith-bank}
of \cref{fig:teaser} against the canonical variable \texttt{(var 0)}.
The body produces an infinite sequence of cells, the first few of
which are summarized in the second column of \cref{fig:replay}.
Now suppose a caller makes the structured call \texttt{(arith-bank
(times a b))}, where \texttt{a} and \texttt{b} are fresh logic
variables in the caller's namespace. The third column of
\cref{fig:replay} shows what the replay does for each canonical
cell: it walks \texttt{(var 0)} in $\sigma_c$, observes the canonical
value (no renaming of internal vars is needed here, since each
pruned canonical cell walks to a ground term), and unifies the
result with \texttt{(times a b)}. Three of the first four cells
fail the head-mismatch check immediately and are dropped, and only
the \texttt{times}-headed cells survive and propagate bindings for
\texttt{a} and \texttt{b} back to the caller.

\begin{figure}[t]
\centering
\small
\begin{tabular}{c@{\hspace{1em}}l@{\hspace{1em}}l}
\toprule
\# & walk*(\texttt{(var 0)},\,$\sigma_c$) & replay against \texttt{(arith-bank (times a b))} \\
\midrule
1 & \texttt{x}              & \texttt{unify (times a b) x}              \,$\to$\, fail \\
2 & \texttt{0}              & \texttt{unify (times a b) 0}              \,$\to$\, fail \\
3 & \texttt{1}              & \texttt{unify (times a b) 1}              \,$\to$\, fail \\
4 & \texttt{(plus x x)}     & \texttt{unify (times a b) (plus x x)}     \,$\to$\, fail \\
5 & \texttt{(times x x)}    & succeed; emit caller cell with \texttt{a $\mapsto$ x, b $\mapsto$ x} \\
6 & \texttt{(plus x 0)}     & \texttt{unify (times a b) (plus x 0)}     \,$\to$\, fail \\
7 & \texttt{(times x 0)}    & succeed; emit caller cell with \texttt{a $\mapsto$ x, b $\mapsto$ 0} \\
\multicolumn{1}{c}{$\vdots$} & \multicolumn{1}{c}{$\vdots$} & \multicolumn{1}{c}{$\vdots$} \\
\bottomrule
\end{tabular}
\caption{A worked walk-through of replay. The middle column shows
the value to which \texttt{(var 0)} walks in the first few canonical
cells of \texttt{arith-bank}; the right column shows what the
replay does for each cell when the caller's argument is the
structured term \texttt{(times a b)}. Only the
\texttt{times}-headed cells survive and add to the caller's
substitution; the rest are silently dropped by unification.}
\label{fig:replay}
\end{figure}

We define \texttt{defrel/memo} helper to enable relations with memoization
against canonical variables and replay.

\subsection{Termination of Recursive Memoized Relations}

It is important to note that a na\"ive ordering would force the body
before installing the cache cell, in which case recursive calls would
re-enter the body and diverge. We avoid this by installing the cache
cell \emph{first} and only then evaluating the body. The recursive
calls inside the body are wrapped by the inverse-$\eta$ delay of
\cref{sec:background-mk}, so that each one is a thunk that fires only
when the answer stream is forced. By the time the stream is forced,
the cache cell is already in place, and the recursive call returns
the canonical stream itself.

\subsection{Memoization With Canonical Pruning}

We now have a single canonical stream that is shared across all call sites,
but it is still unpruned. To get a pruned canonical stream, we simply
wrap the body in \texttt{prune} at canonical-stream construction time.
In the surface syntax, \texttt{defrel/bank}\footnote{We choose
\emph{bank} rather than \emph{tabled} or \emph{memo} to emphasize
that the cached structure is not a tabling table, but a set of
\emph{representatives} pruned by observational equivalence. The
term is borrowed from the non-relational bottom-up PBE literature
(see \cref{sec:related}), where it denotes a worklist of
representative programs grown level by level. Using \emph{tabled}
would mislead, since tabled \miniKanren{} preserves every
syntactically distinct answer (up to variable renaming), while the
bank keeps one representative per \emph{semantic} class; using
\emph{memo} would
underspell, since pure memoization without pruning is the weaker
construct that we call \texttt{defrel/memo}.} is essentially just
\texttt{defrel/memo} plus a \texttt{\#:prune} clause supplying a
deduplication key:

\begin{center}\small
  \texttt{(defrel/bank (rel x \ldots) \#:prune key-expr body \ldots)}.
\end{center}

\noindent The \texttt{key-expr} is evaluated in the scope of the
relation parameters and may therefore refer to them. For our running
example, the key is \texttt{(arith-key e)} from
\cref{sec:background-pbe}. At canonical-stream construction time, the
body is wrapped in \texttt{(prune key-expr \ldots)}, so that the
resulting canonical bank already contains exactly one representative
per behavior, and every replay automatically inherits the
deduplication. In particular, the teaser in \cref{fig:teaser} is
precisely this combination applied to the arithmetic grammar.
Importantly, pruning at the canonical level also sidesteps the
shared-table anti-pattern of \cref{sec:prune-shared}. Indeed, the
canonical input variables \texttt{(var 0)}, \ldots, \texttt{(var
N-1)} are fixed across all call sites, so that ``seen as the outer
\texttt{e}'' and ``seen as the inner \texttt{l}'' are no longer
distinct concepts; they are the same canonical \texttt{(var 0)}.

\subsection{Implementation Notes}
\label{sec:bank-impl}

Two small engineering tricks make the scheme perform competitively
with the depth-bounded baseline of \cref{fig:arith-bounded}. First,
replay walks each input variable in the canonical substitution and
unifies the result, rather than iterating all bindings of the
canonical substitution (most of which are internal-to-internal and
invisible to the caller). This \texttt{walk*}-based replay roughly
halves the cost of replay. Second, when forcing an immature
canonical thunk during replay, we keep forcing through any chain of
nested thunks until a concrete cons cell appears. This is analogous
to \texttt{pull} on the consumer side. Applying forcing on the
producer side as well, we observe that the per-target thunk-force
count drops from $\sim\!330\,000$ to $\sim\!3500$ on the hard
arithmetic targets, bringing \texttt{defrel/bank} within
$1.5\times$ of the depth-bounded baseline before pruning's
algorithmic advantage takes over. We refer the reader to
\cref{sec:appendix-memo} for the full implementation of
\texttt{memo.rkt}; a complete walkthrough is deferred to a future
extended version of this paper.

\section{Best-First Variant: \texttt{defrel/bank-w}}
\label{sec:bankw}

The canonical stream of \texttt{defrel/bank} is built by ordinary
depth-first recursion through \conde{} and \texttt{conj}, so it
emits cells roughly in the order the body explores them. For the
arithmetic grammar of \cref{fig:teaser}, this means that the stream
emits \texttt{x}, \texttt{(times x x)}, \texttt{(times x (times x
x))}, and so on (a long right-spine of \texttt{times} chains) well
before anything starting with \texttt{(plus 1 x)}. Behaviors such as
$(1+x)^k$ then sit far down the bank, and targets that require them
become practically unreachable: the search has to drain thousands of
intervening representatives first. Naturally, increasing the depth
bound does not help, since the bias lies in \emph{which cells the
bank emits first}, rather than in which cells it could in principle
contain.

What we want is a best-first enumeration: emit shallow
representatives before deep ones, regardless of the order the body
happens to walk them. We obtain such an enumeration by attaching
weights to immature stream cells and giving \texttt{mplus} a
sorted-merge discipline.

\subsection{Weights on Immature Streams}

The core data and the \texttt{mplus-w} rule are shown in
\cref{fig:lazy-schema}. The structure \texttt{(lazy weight thunk)}
represents an immature stream paired with an upper bound on the
weight of any cell it could ever emit. The function
\texttt{peek-weight} returns this bound without forcing, and
\texttt{mplus-w} applied to two immature streams returns a single
immature stream whose ceiling is the maximum of the two. Only when
the consumer pulls does \texttt{mplus-w} descend into the side whose
ceiling could still produce the next winning cell. The lazy's
ceiling plays the role of an admissible heuristic in A*-style
search~\cite{hart-nilsson-raphael-1968}: it never under-estimates
the true weight, so a side that loses on peek can be safely
deferred. The design choice of carrying the heuristic on
\emph{immature} stream cells, so that one side can win a comparison
without ever being forced, is inspired by best-first proof-search
tactics, in particular Lean~4's Aesop~\cite{aesop-2023}.

Internally, \texttt{mplus-w} keeps the alternatives of nested merges
in a pairing max-heap keyed by \texttt{peek-weight}, hidden behind
the ordinary stream interface. Thus, nested merges meld their heaps
in constant time, and emitting the next cell costs amortized
logarithmic time in the number of suspended alternatives. This
matters in practice. An earlier version of our implementation built
nests of binary merges, and re-traversing the frontier of suspended
alternatives on every pull dominated the run time of best-first
search (see \cref{sec:eval}).

\begin{figure}[t]
\begin{minted}{racket}
(struct lazy (weight thunk))   ; ceiling on what
                               ; thunk can emit

(define (peek-weight $)
  (cond [(null? $)     -inf.0]
        [(lazy? $)     (lazy-weight $)]
        [else          (car (car $))]))  ; weighted cell

(define (mplus-w $1 $2)
  ;; n-ary sorted merge: alternatives are kept in a
  ;; pairing max-heap keyed by peek-weight; never
  ;; forces a lazy that loses on peek.
  ...)
\end{minted}
\caption{The weighted-stream core. \texttt{lazy} cells advertise an
admissible upper bound, and \texttt{mplus-w} merges them by peek
without forcing.}
\label{fig:lazy-schema}
\end{figure}

\subsection{Why Peek-Without-Force Matters}

It is important that \texttt{peek-weight} does not force its
argument. Indeed, a na\"ive sorted merge that forces both heads in
order to read their weights would re-enter the very canonical stream
that \texttt{defrel/bank-w} is in the middle of building. To see
this, observe that forcing the head of an immature canonical thunk
hits \texttt{memo-thunk}, which calls back into the canonical body,
which recurses into the relation, which asks for the next cell of
the canonical stream, and so on, in an unbounded cycle. The
lazy-ceiling discipline cuts the cycle: the ceiling is fixed at the
moment the \texttt{lazy} is constructed (with \texttt{Zzz-w}
defaulting to $1.0$), and this is enough to order merges without
forcing anything.

\subsection{Depth-Decayed Best-First Enumeration}

The user-facing form, \texttt{defrel/bank-w}\footnote{The
suffix~\texttt{-w} stands for \emph{weighted} and matches our
convention throughout the paper, where weighted analogues of the
standard \miniKanren{} combinators are suffixed with~\texttt{-w}
(\texttt{conde-w}, \texttt{fresh-w}, \texttt{mplus-w}, and so on).},
extends \texttt{defrel/bank} with a decay factor:

\begin{center}\small
  \texttt{(defrel/bank-w (rel x \ldots) \#:prune key-expr \#:decay d body \ldots)}.
\end{center}

\noindent The body uses the weighted versions of standard
combinators (\texttt{conde-w}, \texttt{fresh-w}, and
\texttt{conj-w+}), and the recursive call sites apply
\texttt{(scale-w d)} to their immature streams, multiplying the
ceiling by the decay factor (which defaults to $0.5$). The
exponential depth decay gives shallow representatives the smallest
ceilings, so they emerge from the sorted merge first. With this
discipline, the canonical bank emits \texttt{x}, then \texttt{(plus
x x)}, \texttt{(times x x)}, and \texttt{(plus 1 x)} together at
depth 1, then all the depth-2 representatives, and so on. This is
precisely the order needed to find \texttt{(plus 1 x)$^2$} early.

\subsection{The Trade-Off}

Of course, best-first enumeration finds compact representatives, but
it pays breadth-first costs: emitting any depth-$K$ cell requires
first emitting \emph{all} cells of depth $<K$. For PBE behavior
spaces with thousands of distinct behaviors at depth 3, this is much
slower than \texttt{defrel/bank}'s depth-first drilling. Neither
variant dominates, and \cref{sec:eval} shows that they trade places
across the benchmark suite.

\section{Evaluation}
\label{sec:eval}

The evaluation that follows is deliberately preliminary. Our goal
here is to establish that \texttt{prune} and \texttt{defrel/bank}
yield substantial speedups on a representative slice of PBE
problems, and to surface the regimes in which each variant succeeds
or fails, so that the choice of engine is informed. We make the
limitations of this evaluation explicit at the end of this section,
together with the future evaluations they call for.

The benchmarks we use are drawn from the modules
\texttt{shallow-bench.rkt} (which provides broad coverage of
arithmetic targets at depths 0--3 and string PBE targets at depths
2--3) and \texttt{deep-bench.rkt} (which focuses on deep
arithmetic targets at depths 4--6). Each target consists of three
input/output examples and a minimum search depth, that is, the
smallest depth at which a solution exists. The minimum depth is used
to set the bound for the depth-bounded engine. All measurements are
end-to-end \texttt{run 1} times, that is, the wall-clock time from
query submission to the first matching candidate. The benchmarks
were executed with Racket~9.1~(CS) on an Apple~M4~Pro with 24~GB of
memory, running macOS~26.5. The methodology differs slightly between
the two benchmark scripts. For the shallow targets, each cell is the
mean per-iteration time over 50--1000 iterations (the iteration
count is chosen per target to keep total bench time manageable). For
the deep arithmetic targets, the search times can vary by orders of
magnitude across the engines, so each cell is a single run prefixed
by a forced garbage collection, with per-target timeouts of
30~seconds for targets at depths~$\leq 5$ and 60~seconds for those
at depth~6.

\subsection{Engines}

We compare four engines:
\begin{enumerate}
  \item \texttt{bounded}, the depth-bounded enumerator of
    \cref{fig:arith-bounded}, given the minimum depth for each
    target as $d$.
  \item \texttt{bank}, the \texttt{defrel/bank} of \cref{sec:bank},
    that is, a depth-first canonical enumeration with shared
    deduplication.
  \item \texttt{bank-w}, the \texttt{defrel/bank-w} of
    \cref{sec:bankw} with default decay $0.5$, providing
    depth-decayed best-first enumeration.
  \item \texttt{host-bank}, a non-relational host-language baseline
    (module \texttt{bank.rkt}, \cref{sec:appendix-bank}) that
    builds an exhaustive deduplicated bank in Racket up to the
    target's minimum depth and then exposes its membership predicate
    to a single relational \texttt{run 1} query via \texttt{membero}.
    Unlike \texttt{defrel/bank}, the bank is built outside the
    relational engine and does not compose with arbitrary relational
    goals.
\end{enumerate}

\paragraph{A note on search order.}
The four engines above differ not only in their raw machinery, but
also in the \emph{order} in which they enumerate candidate
programs: \texttt{bounded} follows the natural \conde{} order capped
by an explicit depth bound; \texttt{defrel/bank} emits cells in
canonical recursion order (depth-first, biased toward the right
spine of the \conde{} body); \texttt{defrel/bank-w} uses
depth-decayed best-first ordering; and \texttt{host-bank} enumerates
the bank level by level. Since each measurement is the time to find
the \emph{first} matching candidate, the absolute numbers in
\cref{tab:eval} reflect the interaction between an engine's
enumeration order and where the target's representative happens to
sit in that order, rather than the engines' raw throughput. A
representative that lies early in one engine's enumeration may lie
arbitrarily deep in another's, so the speedup ratios that follow
are best read as characterizing the regimes (target shape
$\times$ enumeration discipline) in which each engine wins, rather
than as universal performance numbers. To make this interaction
explicit, each cell of \cref{tab:eval} is annotated with the
position at which the first matching candidate appears in that
engine's enumeration order.

\begin{table}
  \caption{End-to-end \texttt{run 1} search times, including any bank
    build cost. Bold marks the fastest engine per row; TO denotes a
    timeout (30~seconds for arithmetic targets at depths~$\leq 5$,
    60~seconds for depth~6, and 30~seconds for \texttt{bank-w} on
    string targets). The four engines use different enumeration
    orders (see the discussion above), so the times here measure the
    (target $\times$ order) interaction rather than engine speed in
    isolation. To make the interaction visible, each cell is
    annotated with (\#$n$), the 1-based position of the first
    matching candidate in that engine's enumeration order. The
    indices are deterministic and are produced by the module
    \texttt{order-bench.rkt}.}
  \label{tab:eval}
  \small
  \begin{tabular}{llrrrr}
    \toprule
    Target & Depth & bounded & bank & bank-w & host-bank \\
    \midrule
    \multicolumn{6}{l}{\emph{Arithmetic PBE} (single run; per-target timeout)} \\
    $(1+x)^2$            & 2 & 0.2\,ms\idx{27}           & 0.2\,ms\idx{143}          & 16.4\,ms\idx{29}  & \textbf{$<$0.1\,ms}\idx{25} \\
    $(1+x)^3$            & 3 & 3.3\,ms\idx{128}          & 9.9\,ms\idx{3691}         & 5.28\,s\idx{107}  & \textbf{0.5\,ms}\idx{218} \\
    $x^5$                & 4 & 0.6\,ms\idx{44}           & \textbf{0.1\,ms}\idx{36}  & 126\,ms\idx{36}   & 164\,ms\idx{368}          \\
    $(1+x)^4$            & 4 & \textbf{139\,ms}\idx{1061} & 1.99\,s\idx{243158}      & TO       & 172\,ms\idx{182}          \\
    $x^5 + x$            & 5 & 3.3\,ms\idx{88}           & \textbf{0.1\,ms}\idx{71}  & 14.7\,s\idx{174}  & TO               \\
    $x^5 + 1$            & 5 & 24.3\,ms\idx{165}         & \textbf{0.7\,ms}\idx{302} & 26.4\,s\idx{202}  & TO               \\
    $x^6$                & 5 & 3.6\,ms\idx{89}           & \textbf{0.2\,ms}\idx{72}  & 12.6\,s\idx{169}  & TO               \\
    $(1+x)^5$            & 5 & \textbf{6.58\,s}\idx{10055} & TO                      & TO       & TO               \\
    $x^6 + 1$            & 6 & 139\,ms\idx{339}          & \textbf{1.4\,ms}\idx{594} & TO       & TO               \\
    $x^7$                & 6 & 20.1\,ms\idx{181}         & \textbf{0.3\,ms}\idx{145} & TO       & TO               \\
    \midrule
    \multicolumn{6}{l}{\emph{String PBE} (mean per iteration over 10--200 iterations)} \\
    \texttt{"Hello, "$+$X$+$"!"}   & 2 & \textbf{0.61\,ms}\idx{349}  & 2.92\,ms\idx{1367}  & 1.18\,ms\idx{507} & 3.69\,ms\idx{3023} \\
    \texttt{"Hi, "$+$X$+$"?"}      & 2 & \textbf{0.72\,ms}\idx{416}  & 7.38\,ms\idx{2946}  & 1.22\,ms\idx{517} & 3.74\,ms\idx{3598} \\
    \texttt{X$+$", "$+$X}          & 2 & \textbf{0.09\,ms}\idx{59}   & 0.32\,ms\idx{221}   & 0.25\,ms\idx{98}  & 3.19\,ms\idx{433} \\
    \texttt{"Hello, "$+$X$+$"!!"}  & 3 & 31.5\,ms\idx{7131} & 30.3\,ms\idx{9505}  & \textbf{10.4\,ms}\idx{1212} & 123\,s\idx{2972}   \\
    \bottomrule
  \end{tabular}
\end{table}

\subsection{Discussion of the Results}

We now discuss the results in \cref{tab:eval} engine by engine.

\paragraph{\texttt{bounded}: predictable, but rate-limited by depth.}
The depth-bounded enumerator is the only engine that never times out
on the deep arithmetic targets, including the hard $(1+x)^k$ row.
Indeed, pruning at every recursive level is correct, and the cheap
depth cut keeps the candidate frontier bounded. The cost, however,
shows up in absolute terms: at depths 5 and 6, \texttt{bounded} is
consistently one to three orders of magnitude slower than
\texttt{bank} on the targets they both solve. This is because every
call site allocates a fresh deduplication hash and re-prunes the
same sub-expressions that the previous level has already pruned.

\paragraph{\texttt{bank}: roughly 9--99$\times$ faster on 6 out of 8
deep arithmetic targets, except the $(1+x)^k$ family.}
On the deep arithmetic suite, \texttt{defrel/bank} is between roughly
$9\times$ and $99\times$ faster than \texttt{bounded} on every target
whose representative lies along the right-spine of the \conde{} body,
that is, the $x^k$ and $x^k + c$ family. We stress, however, that
these ratios are a property of the (target, enumeration-order)
pair rather than of \texttt{defrel/bank} itself, since
\texttt{defrel/bank} and \texttt{bounded} explore the candidate
space in different orders. For the right-spine family, the bank's
canonical enumeration places the answer very early; for the
$(1+x)^k$ family of the same depth, the same canonical enumeration
places the answer very late, and the ratio inverts. Specifically,
the bank either loses by a wide margin (at depth~4, $(1+x)^4$ is
solved by the bank in $1.99$~seconds versus $139$~ms for
\texttt{bounded}) or times out altogether (at depth~5, $(1+x)^5$ is
found by \texttt{bounded} in $6.58$~s while both the bank and the
weighted variant time out).
The reason is that the canonical bank emits right-spine
\texttt{times} chains long before anything starting with
\texttt{(plus 1 x)}. Indeed, the representative of $(1+x)^2$ sits
at position 143 in the bank, that of $(1+x)^3$ at position 3691,
and that of $(1+x)^4$ at position 243\,158 (see the indices in
\cref{tab:eval}). We have also verified
that fair conjunction in the style of Kiselyov, Shan, Friedman, and
Sabry~\cite{kiselyov-shan-friedman-sabry-2005} does \emph{not} fix
this: an experimental \texttt{conj-i} variant
(\texttt{mplus-i}/\texttt{bind-i} with diagonal pairing) reproduces
the same first 30 canonical cells as the standard \texttt{conj}.
The bias lies in \emph{which cells exist} in the bank, not in
\emph{how pairs of cells are formed}, which is a problem fair
conjunction cannot address.

\paragraph{\texttt{bank-w}: compact representatives, competitive on
wide-but-shallow behavior spaces.}
The weighted variant \texttt{defrel/bank-w} reverses the trade-off.
Whenever it terminates, it returns the most compact representative
of the target behavior. For example, for $(1+x)^3$ it finds
\texttt{(times (times (plus 1 x) (plus 1 x)) (plus 1 x))}, where
\texttt{bank} returns a sprawling depth-8 term. Its enumeration indices are
correspondingly small (\#29 for $(1+x)^2$ and \#107 for $(1+x)^3$,
versus \#143 and \#3691 for \texttt{bank}). On the string suite,
\texttt{bank-w} is competitive across the board and wins the
depth-3 row outright ($10.4$\,ms versus $31.5$\,ms for
\texttt{bounded}). On deep arithmetic, however, emitting any
depth-$K$ cell still requires emitting every lighter cell first,
and the four hardest rows time out. We see \texttt{bank-w} as the
right tool when the behavior space is wide but the target is
shallow, and when obtaining the compact representative matters.

Our original expectation for \texttt{bank-w} was more optimistic.
We hoped that best-first enumeration would neutralize the
enumeration-order bias of \texttt{defrel/bank} at an acceptable
cost, making \texttt{bank-w} the default engine. In the version of
this paper submitted for review, this expectation failed badly.
There, \texttt{bank-w} timed out on all arithmetic targets except
$(1+x)^2$ and $x^5$, and also on the depth-3 string target. A
reviewer asked us to explain why. Investigating the question
revealed an implementation artifact rather than a fundamental
limit. The enumeration indices show that the number of
\emph{emitted} cells is modest, since pruning keeps one
representative per behavior. However, our original \texttt{mplus-w}
built nests of binary merges, so emitting each cell re-traversed
the entire frontier of suspended alternatives. For example, on
$(1+x)^2$, emitting 34 cells cost $66$\,ms, roughly $2$\,ms per
cell.\footnote{The count differs from the (\#29) reported in
\cref{tab:eval} because the pairing-heap merge breaks ties within
an equal-weight class differently than the binary nests did. Both
orders are deterministic.} This is three orders of magnitude more than \texttt{bank}'s
per-cell cost on the same target. Replacing the nests with the
pairing-heap merge of \cref{sec:bankw} improved \texttt{bank-w} by
one to three orders of magnitude across the suite: $(1+x)^2$ from
$66$\,ms to $16$\,ms, $x^5$ from $4.1$\,s to $126$\,ms, $(1+x)^3$
and three depth-5 targets from timeout to seconds, and the depth-3
string target from timeout to $10.4$\,ms. \Cref{tab:eval} reports
the heap-based numbers. What remains is the genuine breadth-first
cost. Every cell of a weight class must be emitted before any cell
of a lower class, and the class sizes grow with the number of
distinct behaviors per level (already thousands at depth~3 for
arithmetic). This is why the deepest arithmetic rows still time
out.

\paragraph{The decay factor.}
A natural question is how sensitive \texttt{bank-w} is to the value
of the decay factor $d$. In exact arithmetic, it is not sensitive
at all. With a uniform decay applied at every recursive call, the
weight of every canonical cell is $d^n$, where $n$ counts the decay
applications in the cell's derivation. For every $d \in (0,1)$,
$d^n$ is monotone in $n$. Thus, the pairwise comparisons made by
\texttt{mplus-w} reduce to comparisons of derivation sizes, and the
enumeration order does not depend on $d$. In floating-point
arithmetic, the picture is subtler. We verified it empirically for
$d \in \{0.9, 0.75, 0.5, 0.25, 0.1\}$ (module
\texttt{decay-bench.rkt}). The sequence of \emph{weight classes} in
the first 50 canonical cells is identical for all five values.
However, the order \emph{within} an equal-weight class is perturbed
for $d = 0.9$ and $d = 0.1$, whose powers are not exactly
representable in double precision. Indeed, the observed divergence
points sit exactly at class boundaries. The values $d \in \{0.75,
0.5, 0.25\}$, whose powers are exactly representable at these
sizes, produce literally identical orders. The within-class order
can matter in practice. For example, on the $(1+x)^2$ target, the
accidental order induced by $d = 0.9$ finds the target roughly
$6\times$ faster than the exact-arithmetic order ($2.8$\,ms versus
$17$\,ms). This is luck rather than principle. Thus, the decay knob
becomes meaningful only once different productions carry different
weights (e.g. Probe-style learned weights~\cite{probe-2020}), which
we leave to future work.

\paragraph{\texttt{host-bank}: complementary strengths, but
non-compositional.}
The non-relational \texttt{host-bank} baseline exhibits a behavior
that is, in many ways, complementary to \texttt{defrel/bank}. On
shallow targets, the host bank is the fastest engine. In particular,
on the $(1+x)^k$ family for $k \in \{2, 3\}$, \texttt{host-bank}
finds the compact representative in fractions of a millisecond,
where \texttt{defrel/bank} either ties or loses by an order of
magnitude. At depth~4, \texttt{host-bank} solves $(1+x)^4$ in
$165$~ms, comparable to \texttt{bounded}'s $141$~ms and an order of
magnitude better than \texttt{defrel/bank}'s $1.99$~s. The reason
is that \texttt{host-bank}'s level-by-level construction does not
suffer from the right-spine bias of \cref{sec:bankw}, and it sees
$(1+x)$ as early as any other depth-1 representative.

On the deep right-spine targets, however, \texttt{host-bank} loses
catastrophically. Already at $x^5$ (depth~4) it spends $154$~ms,
two orders of magnitude slower than the $0.1$~ms of
\texttt{defrel/bank}. From depth~5 onwards, the cost of building the
exhaustive bank exceeds the search budget and \texttt{host-bank}
times out on every target. The same pattern shows up on the string
targets, where the host bank is competitive with \texttt{defrel/bank}
at depth~2 but degrades to $119$~s per iteration at depth~3, as the
bank build has to enumerate every depth-3 string composition.

In any case, the host bank achieves whatever performance it has by
abandoning the relational interface: the bank is built outside
\miniKanren{}, and the membership predicate cannot be combined with
arbitrary relational goals such as typed enumerators, refinement
constraints, or mutual recursion across relations. By contrast,
\texttt{defrel/bank} keeps the bank inside the relational language,
recovering most of the host-bank speedup where the right-spine bias
does not bite, and remaining composable with the rest of
\miniKanren{}.

\paragraph{Which engine, when.}
The preceding paragraphs suggest simple selection guidance. Use
\texttt{bounded} when the target depth is known (or can be iterated
over) and predictability matters most. It is the only engine that
never times out in our suite. It also remains the fastest on the
shallow string rows and on deep targets whose representatives sit
late in the canonical order (the $(1+x)^k$ family). Use
\texttt{defrel/bank} as the default depth-less engine. It wins by
one to two orders of magnitude whenever the target's representative
sits early in the canonical order (the $x^k$ and $x^k + c$ family).
Its failure mode is confined to representatives that sit late in
that order. Use \texttt{defrel/bank-w} when the behavior space is
wide but the target is shallow (it wins the depth-3 string row), or
when obtaining the compact representative matters more than raw
speed. Finally, use the host-language bank when compositionality
with other relational goals is not needed and the minimum depth is
small.

\paragraph{Limitations of this evaluation.}
We close the section by making the limitations of this evaluation
explicit.
\begin{itemize}
  \item The suite covers two domains with one grammar each, 14
    targets in total. Conclusions about which enumeration order wins
    on which target shape may not transfer to richer grammars.
  \item The deep arithmetic cells are single runs. Ratios drift
    across executions, and cells close to the budget may flip to a
    timeout. For example, \texttt{bank-w} solves $x^5 + 1$ in
    $26.4$\,s against a $30$\,s budget.
  \item \texttt{bounded} is given the minimum depth for each target,
    which is its best case. In practice the right depth is unknown,
    and the cost of discovering it (e.g. by iterative deepening) is
    not measured.
  \item All four engines are our own implementations over the same
    core. We do not yet compare against tabled \miniKanren{},
    relational-interpreter synthesis, or bottom-up SyGuS solvers.
  \item We measure wall-clock time to the first answer only. Memory
    consumption (the canonical bank persists for the whole run) is
    not reported.
\end{itemize}
Correspondingly, we plan the following evaluations for an extended
version of this paper, each matched to an open question:
\begin{itemize}
  \item additional PBE domains (bit-vector, list, typed-component
    synthesis), to test whether the enumeration-order regimes of
    \cref{tab:eval} persist on richer grammars;
  \item a comparison with tabled \miniKanren{} on relations where
    both apply, to quantify the cost of semantic versus syntactic
    deduplication;
  \item a comparison with bottom-up SyGuS solvers on a common subset
    of targets, to locate the overhead of staying relational;
  \item repeated-trial timing with memory profiling.
\end{itemize}

\section{Related Work}
\label{sec:related}

We organize the related work into four threads, covering bottom-up
PBE synthesis, prior work on synthesis via \miniKanren{}, tabling
in logic programming, and fair search.

\subsection{Bottom-Up PBE Synthesis}

Bottom-up enumeration with observational deduplication is the
central idea behind a long line of non-relational PBE
synthesizers~\cite{albarghouthi-2013,eusolver-2017,probe-2020,bustle-2021}.\footnote{We
use ``relational'' here in the \miniKanren{} sense, that is, programs
written as multi-way relations over logic variables. Wang, Wang, and
Dillig~\cite{wang-wang-dillig-2018} use the same word for the
unrelated notion of synthesizing pairs of programs satisfying a
relational specification.}
These tools maintain a worklist of representative expressions,
deduplicated by their behavior on the example inputs, and grow the
worklist level by level until a candidate matches the
specification. We adopt the same bottom-up plus
observational-equivalence skeleton in \texttt{defrel/bank}: the
canonical-variable bank \emph{is} a bottom-up worklist, except
that it is consumed by \miniKanren{}'s unification-driven search
rather than by an explicit enumeration loop, and it composes with
arbitrary relational goals such as typed enumerators, refinement
constraints, or mutual recursion across relations. One specific
neighbor worth singling out, on the \texttt{defrel/bank-w} side
rather than the \texttt{defrel/bank} side, is Barke, Peleg, and
Polikarpova's Probe~\cite{probe-2020}, which layers a
just-in-time-learned probabilistic context-free grammar over a
bottom-up behavior-pruned bank to bias the enumeration order toward
likely solutions. The closest analog in our paper is the much
simpler depth-decay heuristic of \texttt{defrel/bank-w}
(\cref{sec:bankw}); layering a Probe-style learned ranker on top of
\texttt{defrel/bank-w} would be a natural direction for future work. A complementary angle on the
same goal of shrinking the effective PBE search space is taken by
Hocquette and Cropper~\cite{hocquette-cropper-2025}, who decompose
each example into position-indexed input/output facts and learn
relations between those facts via inductive logic programming.
Their decomposition acts on the \emph{representation} side,
reshaping how the examples are presented to the synthesizer; our
\texttt{prune} acts on the \emph{enumeration} side, quotienting
candidate answers by observational equivalence.

\subsection{Synthesis via \miniKanren{}}

The closest prior work on \emph{relational} synthesis uses
\miniKanren{} as a top-down search engine over relational
interpreters. The idea is to run a relational evaluator
``backwards'' against the target input/output examples, and the
search then returns programs whose evaluation matches the examples.
Byrd, Ballantyne, Rosenblatt, and
Might~\cite{byrd-ballantyne-rosenblatt-might-2017} demonstrate this
recipe on a suite of synthesis problems, and the Barliman
prototype~\cite{barliman} packages it into a live program-completion
tool. Hemann and Friedman~\cite{hemann-friedman-2020} extend the
canon with further quine-style benchmarks and ``mirrored''
relational-interpreter tasks. The same recipe has also been applied
to several non-toy languages: Chirkov \emph{et
al.}~\cite{chirkov-rosenblatt-might-zhang-2020} build a relational
interpreter for a subset of JavaScript by composing a relational
S-expression parser with a relational evaluator, Kosarev, Lozov, and
Boulytchev~\cite{kosarev-lozov-boulytchev-2020} synthesize
pattern-matching decision trees by running a low-level switch
relation backward against a high-level pattern-match relation, and
Domoratskiy and
Boulytchev~\cite{domoratskiy-boulytchev-2024} report the OCanren
extensions and optimizations needed to scale a relational
type-inference solver beyond toy STLC examples. The
\texttt{prune} and \texttt{defrel/bank} combinators of the present
paper are in the same spirit of small additions to vanilla
\miniKanren{} that aim to make this recipe scale.

Two recent directions are worth contrasting more closely with our
work. First, Rosenblatt, Zhang, Byrd, and
Might~\cite{rosenblatt-zhang-byrd-might-2019} introduce a first-order
defunctionalized representation of \miniKanren{} goals and streams
that decouples search from semantics, making it easier to swap in
alternate search strategies, and Zhang \emph{et
al.}~\cite{neuralkanren-2018} pursue one such alternate strategy,
training a neural network to score candidate branches in the
\miniKanren{} search tree and thereby guide the search toward
promising sub-searches on PBE problems. Our emphasis is different:
we focus on memoization of canonical-variable streams and on
pruning those streams by observational equivalence, rather than on
the order in which branches of a top-down search are expanded.
Second, Ballantyne \emph{et al.}~\cite{ballantyne-multistage-2025}
attack the per-query overhead of synthesis-via-relational-interpreters
from a different angle, lifting MetaOCaml-style multi-stage
programming into \miniKanren{} so that the known parts of a
partially-unknown program can be compiled away, leaving only the
relational holes to be solved. Their staging and our memoization are
complementary: their technique cuts the per-step cost of the
relational interpreter, whereas \texttt{defrel/bank} cuts the cost
of re-enumerating the same answer set across recursive call sites of
a sub-relation.

We view all of these efforts as orthogonal to \texttt{defrel/bank}:
relational interpretation describes the synthesis problem
declaratively, while \texttt{defrel/bank} adds bottom-up
enumeration to the underlying search, and we believe that the
several lines compose. The branch-selection question of Zhang
\emph{et al.}~\cite{neuralkanren-2018} does surface narrowly for
our weighted variant \texttt{defrel/bank-w}, where we answer it
with an admissible-heuristic ceiling on immature streams rather
than with a learned scorer; combining the two answers would be an
interesting direction for future work.

\subsection{Tabling in Logic Programming}

Tabling, that is, memoizing the answer set of a relation, has a long
history. The foundational technique is OLD resolution with
tabulation~\cite{tamaki-sato-1986}, and SLG resolution as
implemented in XSB~\cite{chen-warren-1996} extends it to general
logic programs with negation. Tabled \miniKanren{} is the
descendant of these ideas in the relational
host~\cite[Part~IV]{byrd-thesis-2009}, and it is the closest
neighbour to \texttt{defrel/bank}. Indeed, both systems memoize a
relation on its first call and replay the cached answers on
subsequent calls, and both must handle the variable-shifting issue
that arises when canonical bindings are substituted into the
caller's namespace.

That said, there are two differences, and both are best understood
by viewing \texttt{defrel/bank} as a variant of tabling rather than
as an unrelated construct. The first difference is \emph{which
equivalence} the cache preserves. Classical tabling already operates
on equivalence classes of answers, namely syntactic identity up to
variable renaming. \texttt{defrel/bank} swaps this fixed syntactic
equivalence for a user-supplied semantic one (in our PBE setting,
behavior on the example inputs). For PBE, the semantic classes are
the desired shape, since two terms with the same input/output
behavior are interchangeable for the rest of the search. The price
is that the cached stream no longer reproduces the full syntactic
answer set. This matters for multiplicity-sensitive applications.
In probabilistic logic programming, for instance, the number of
syntactically distinct answers may itself be the quantity of
interest, and collapsing by behavior would be unsound there.

The second difference lies in cache indexing. Tabled \miniKanren{}
keys each cache by the reified call arguments, so that different
call shapes maintain separate caches. Semantic deduplication, if
added, would then have to be repeated per cache. By contrast,
\texttt{defrel/bank} fixes the cache key to the most general call.
The single canonical bank is built with all parameters unbound, and
caller-specific structure is recovered by unification at replay
time. In this light, \texttt{defrel/bank} may fairly be described
as tabling with the argument key fixed to the most general call,
plus deduplication by a user-chosen key. Indeed, one could
approximate it in a classical tabling system by always calling the
tabled relation with unbound arguments and constraining the results
afterwards. What \texttt{defrel/bank} adds on top of this recipe is
that deduplication happens once, inside the shared cache, rather
than downstream of every call site. These trade-offs add up to a difference
in implementation cost as well: tabled \miniKanren{} extends the
central stream dispatcher \texttt{case-inf} with a new waiting-stream
variant and rewrites \texttt{take}, \texttt{bind}, and
\texttt{mplus} to detect saturation, whereas \texttt{defrel/bank}
sits as a library on top of an unmodified \muKanren{} core,
requiring only a per-\texttt{(run \ldots)} parameter for the cache.
We see the two designs as complementary, and a future extension may
thread a tabling mode through \texttt{defrel/memo}, keyed on
whether the caller wants every answer or every equivalence class.

\subsection{Fair Search and Search Strategies}

Fair conjunction and disjunction ensure that every answer is
eventually emitted, by interleaving the streams from each conjunct
or disjunct. The monad-transformer view is due to
Hinze~\cite{hinze-2000} and Kiselyov, Shan, Friedman, and
Sabry~\cite{kiselyov-shan-friedman-sabry-2005}, while the algebraic
view of Spivey~\cite{spivey-2009} characterizes depth-first,
breadth-first, and iterative-deepening search as instances of a
single search algebra. \miniKanren{}'s standard \texttt{mplus}
implements interleaved disjunction, and the corresponding
fair-conjunction discipline has been studied specifically for
\miniKanren{} by Lozov and Boulytchev~\cite{lozov-boulytchev-2020},
whose fair-conjunction combinator converges independently of
conjunct order, and by Lu, Ma, and
Friedman~\cite{lu-ma-friedman-2019}, who survey fair-search
strategies in the relational setting. Rozplokhas and
Boulytchev have studied the same question from two further angles:
in earlier work~\cite{rozplokhas-boulytchev-2018}, they introduce a
dynamic divergence test that detects potentially non-terminating
conjuncts at run time and reorders them to improve refutational
completeness; in later work~\cite{rozplokhas-boulytchev-2022}, they
analyze the scheduling complexity of interleaving search,
characterizing when fair disciplines yield asymptotic improvements.
The closest
\miniKanren{}-internal predecessor to a weighted-stream discipline
such as ours is Swords and Friedman's
\texttt{rKanren}~\cite{swords-friedman-2013}, which introduces
uniform-cost guided search inside \miniKanren{}.

As we have discussed in \cref{sec:eval}, the depth-bias of
\texttt{defrel/bank} on $(1+x)^k$ is not a fair-conjunction failure:
the bias is in \emph{which cells} the canonical stream emits first,
not in \emph{how pairs of cells} are formed. The lazy-ceiling
discipline of \texttt{defrel/bank-w} lives in the same design space
as these fair-search efforts, but it is closer to a relational
analogue of A* (a sorted merge by an admissible heuristic on
immature streams) than to interleaving or uniform-cost search.

\section{Conclusion and Future Work}
\label{sec:conclusion}

We have presented two small library combinators, \texttt{prune} and
\texttt{defrel/bank}, together with a weighted variant
\texttt{defrel/bank-w}, that suffice to lift bottom-up
enumeration with observational deduplication into \miniKanren{}
without changing the host search discipline. On a preliminary PBE
benchmark of arithmetic and string synthesis targets,
\texttt{defrel/bank} is 9--99$\times$ faster than the depth-bounded
baseline on 6 out of 8 deep arithmetic targets, and the two cases
it loses, both in the $(1+x)^k$ family, are precisely characterized
as an enumeration-order issue that fair conjunction cannot resolve.
We plan a more thorough empirical study for an extended version of
this paper.

We see several promising directions for future work.

First, the depth bias of \texttt{defrel/bank} on $(1+x)^k$ may be
addressed by \emph{iterative deepening over the canonical body},
that is, by forcing the canonical bank to enumerate depth $K$ fully
before any depth $K+1$, perhaps by instrumenting the recursive calls
with a depth counter. We expect that such a discipline would fix the
$(1+x)^k$ bias by construction, without paying the breadth-first
cost that \texttt{defrel/bank-w} pays on wide behavior spaces.

Second, an obvious optimization is \emph{saturation detection}:
stop growing the canonical bank once the prune cache stops
expanding. This caps the cost of unhelpful bank construction, and is
also a prerequisite for the tabling-mode semantics that we have
sketched in \cref{sec:related}.

Third, we may consider \emph{mixed depth-first and best-first
strategies}, for example, preferring depth-first within a sub-tree
and best-first across sub-trees. One way to achieve this is to allow
\texttt{scale-w} to be applied selectively, or to combine
\texttt{mplus-w} and \texttt{mplus} at different recursion boundaries.

Fourth, it remains to be seen how our combinators interact with the
relational-interpreter
recipe~\cite{byrd-ballantyne-rosenblatt-might-2017} and, by
extension, with systems such as Barliman~\cite{barliman}, where the
examples are test cases for a program executed by a relational
interpreter. The enumerate-and-test recipe that we target in this
paper keeps the candidate generator separate from the oracle. This
is what lets \texttt{ground-key} evaluate candidates with a
host-language interpreter. It is an open question whether a
behavioral prune key can be computed efficiently for candidates
\emph{inside} a relational interpreter for a realistic language,
e.g. by running each candidate on the test inputs as part of the
key. We plan to investigate this.

Finally, the preliminary benchmark of \cref{sec:eval} should be
broadened along the lines set out in the limitations paragraph at
the end of that section, and the interaction of our combinators
with the existing \miniKanren{} constraint stores remains to be
studied.

\begin{acks}
  We thank the anonymous miniKanren 2026 reviewers for their
  helpful comments, and in particular for the question about
  \texttt{defrel/bank-w}'s performance that prompted the
  pairing-heap merge of \cref{sec:bankw}.
\end{acks}

\bibliographystyle{ACM-Reference-Format}
\bibliography{ms}

\appendix

\section{Supplementary Material}
\label{sec:appendix-supp}

In this appendix, we provide the full source code of our prototype
implementation, together with the setup instructions and the
reproduction steps for the numerical results reported in
\cref{tab:eval}. The prototype is implemented in
Racket and consists of twelve modules totalling roughly 2000 lines
of code, all of which we list in
\cref{sec:appendix-source}. The modules keep the layout of the
accompanying repository: the library at the root, the benchmark
drivers under \texttt{bench/}, and the shared-table witness under
\texttt{examples/}.

\subsection{Overview}
\label{sec:appendix-overview}

\Cref{tab:supp-files} provides an overview of the modules in the
prototype. For each module, we indicate the section of the main
text that introduces or makes use of it, together with a pointer to
the corresponding source listing later in this appendix.

\begin{table}[h]
  \caption{Modules in the appendix.}
  \label{tab:supp-files}
  \small
  \begin{tabular}{lp{5.2cm}ll}
    \toprule
    Module & Description & Paper~\S & Listing \\
    \midrule
    \texttt{microkanren.rkt}   & Minimal \muKanren{} core
                                 plus fair and weighted stream primitives
                               & \S\ref{sec:background-mk}             & \S\ref{sec:appendix-microkanren} \\
    \texttt{wrappers.rkt}      & Variadic and weighted surface wrappers
                                 (\texttt{conde}, \texttt{fresh}, \texttt{run},
                                 \texttt{conde-w}, \ldots)
                               & \S\ref{sec:background-mk}             & \S\ref{sec:appendix-wrappers} \\
    \texttt{prune.rkt}         & The \texttt{prune} combinator and helpers
                               & \S\ref{sec:prune}                     & \S\ref{sec:appendix-prune} \\
    \texttt{memo.rkt}          & \texttt{defrel/memo},
                                 \texttt{defrel/bank},
                                 \texttt{defrel/bank-w}
                               & \S\ref{sec:bank},\,\S\ref{sec:bankw}  & \S\ref{sec:appendix-memo} \\
    \texttt{bench/bank.rkt}    & Non-relational host-language bank
                                 (the \texttt{host-bank} engine) and
                                 \texttt{membero}
                               & \S\ref{sec:eval}                      & \S\ref{sec:appendix-bank} \\
    \texttt{main.rkt}          & Re-export hub for the public API
                               & ---                                   & \S\ref{sec:appendix-main} \\
    \texttt{info.rkt}          & Racket collection metadata
                               & ---                                   & \S\ref{sec:appendix-info} \\
    \texttt{examples/shared-table-witness.rkt}  & Witness for the shared-table
                                 anti-pattern (completeness failure)
                               & \S\ref{sec:prune-shared}              & \S\ref{sec:appendix-witness-listing} \\
    \texttt{bench/shallow-bench.rkt} & Shallow PBE benchmark driver:
                                 arith depths 0--3 and string PBE
                                 depths 2--3
                               & \S\ref{sec:eval}                      & \S\ref{sec:appendix-shallow-bench} \\
    \texttt{bench/deep-bench.rkt}    & Deep arithmetic PBE benchmark driver:
                                 depths 4--6, four engines
                               & \S\ref{sec:eval}                      & \S\ref{sec:appendix-deep-bench} \\
    \texttt{bench/order-bench.rkt}   & First-answer enumeration indices
                                 per engine and target (the (\#$n$)
                                 annotations of \cref{tab:eval})
                               & \S\ref{sec:eval}                      & \S\ref{sec:appendix-order-bench} \\
    \texttt{bench/decay-bench.rkt}   & Decay-factor experiment for
                                 \texttt{defrel/bank-w}
                               & \S\ref{sec:eval}                      & \S\ref{sec:appendix-decay-bench} \\
    \bottomrule
  \end{tabular}
\end{table}

\subsection{Setup}
\label{sec:appendix-setup}

The prototype is implemented in pure Racket and runs on Racket~9.x
(CS) on macOS, Linux, or Windows; the numbers reported in this paper
were produced with Racket~9.1~(CS). No external packages are
required beyond the Racket \texttt{base} collection, and the
installation instructions for Racket itself may be found at
\url{https://racket-lang.org/}.

Once the source tree is unpacked (or the repository at
\url{https://github.com/fizruk/prune-kanren} is cloned), the modules
may be optionally pre-compiled for faster benchmark startup with

\begin{center}\small\texttt{raco make main.rkt bench/*.rkt examples/*.rkt}\end{center}

\noindent We also provide the package metadata in \texttt{info.rkt}
(\cref{sec:appendix-info}), which allows the directory to be picked
up by \texttt{raco~pkg} as a local Racket package, should the user
prefer to install it.

\subsection{Reproducing \cref{tab:eval}}
\label{sec:appendix-repro}

The two row groups of \cref{tab:eval} are produced by two
benchmark drivers, which we describe in turn.

\paragraph{Deep arithmetic rows.}
To produce the deep arithmetic rows of \cref{tab:eval}, the reader
may run

\begin{center}\small\texttt{racket bench/deep-bench.rkt}\end{center}

\noindent from the root of the source tree. The
script runs ten arithmetic targets at depths 2 through 6 against
each of the four engines (\texttt{bounded}, \texttt{bank},
\texttt{bank-w}, \texttt{host-bank}), with per-target wall-clock
timeouts of 30~seconds for depths $\leq 5$ and 60~seconds for
depth~6. Each cell is the wall-clock time of a single run, prefixed
by a forced garbage collection; the full source is in
\cref{sec:appendix-deep-bench}. We expect the total runtime to be
roughly 3 to 6~minutes on a modern laptop, dominated by the rows
on which multiple engines reach the timeout. Because the
methodology is single-run, the exact ratios may drift slightly
across executions; for instance, we have observed the $x^5$ row's
bank-versus-\texttt{bounded} ratio moving between roughly
$9\times$ and $11\times$ across two runs on the same hardware. The
qualitative shape of the table, however, is stable.

\paragraph{String PBE rows.}
To produce the string PBE rows of \cref{tab:eval}, the reader may
run

\begin{center}\small\texttt{racket bench/shallow-bench.rkt}\end{center}

\noindent from the same directory. This second driver covers both
the arithmetic suite at depths 0--3 (seven targets, mean per
iteration over 50--1000 iterations) and the string PBE suite at
depths 2--3 (four targets, mean per iteration over 10--200
iterations). For each string target, all four engines are timed;
\texttt{bank-w} is additionally wrapped in a 30~second timeout,
since it can be slow on wide string-behavior spaces. The full
source is in \cref{sec:appendix-shallow-bench}. We expect the total
runtime to be roughly 20 to 25~minutes, almost all of it spent on
the depth-3 string target with \texttt{host-bank}, which rebuilds
the exhaustive depth-3 string bank on every iteration.

\paragraph{Enumeration indices.}
The (\#$n$) annotations of \cref{tab:eval} are produced by

\begin{center}\small\texttt{racket bench/order-bench.rkt}\end{center}

\noindent which counts, for each engine and target, the candidates
emitted up to and including the first match. Unlike the timings,
these indices are deterministic and machine-independent. The run
takes roughly 15~minutes, dominated by the cells that reach their
timeout while counting.

\paragraph{Decay-factor experiment.}
The numbers in the decay-factor discussion of \cref{sec:eval} are
produced by

\begin{center}\small\texttt{racket bench/decay-bench.rkt}\end{center}

\noindent which enumerates the first 50 representatives of the
weighted arithmetic bank for five decay values, compares the
prefixes and their weight-class structure, and times the $(1+x)^2$
target under each value. The run takes well under a minute.

\paragraph{Hardware.}
The numbers reported in this paper were measured on an
Apple~M4~Pro with 24~GB of memory, running macOS~26.5 and
Racket~9.1~(CS).

\subsection{Verifying the Shared-Table Anti-Pattern of
\cref{sec:prune-shared}}
\label{sec:appendix-witness}

To reproduce, in practice, the completeness failure that we discuss
in \cref{sec:prune-shared}, the reader may run

\begin{center}\small\texttt{racket examples/shared-table-witness.rkt}\end{center}

\noindent from the same directory. This script defines a variant of
\texttt{prune} that threads a single deduplication table across all
nested calls via a Racket parameter, and runs it side by side with
the standard per-call variant on two PBE targets. On the easy
target $x \cdot x$, the shared variant hangs: the depth-less search
loops looking for an alternative ground term with behavior $(4, 9,
16)$, finds none, and never returns. By contrast, the standard
per-call \texttt{prune} variant solves the same target immediately.
The full source is given in \cref{sec:appendix-witness-listing}; the
run should be interrupted with Ctrl-C once the failure mode is
visible.

\subsection{Source Code}
\label{sec:appendix-source}

The remainder of this appendix lists the full source of each module
of the prototype in dependency order. We begin with the
\muKanren{} core (\cref{sec:appendix-microkanren}) and the surface
wrappers around it (\cref{sec:appendix-wrappers}), which together
constitute the substrate that we build upon. We then list the new
combinators introduced in this paper, namely \texttt{prune} and its
helpers (\cref{sec:appendix-prune}) and the memoization combinators
\texttt{defrel/memo}, \texttt{defrel/bank}, and
\texttt{defrel/bank-w} (\cref{sec:appendix-memo}). Next, we provide
the non-relational host-language bank used as the \texttt{host-bank}
engine in \cref{sec:eval} (\cref{sec:appendix-bank}), followed by
the package files \texttt{main.rkt} and \texttt{info.rkt}
(\cref{sec:appendix-main,sec:appendix-info}). Finally, we list the
witness for the shared-table anti-pattern and the four benchmark
drivers that produce the numbers reported in this paper
(\cref{sec:appendix-witness-listing,sec:appendix-shallow-bench,sec:appendix-deep-bench,sec:appendix-order-bench,sec:appendix-decay-bench}).

\subsubsection{\texttt{microkanren.rkt}}
\label{sec:appendix-microkanren}

\begin{minted}{racket}
#lang racket/base

;; Minimal microKanren core, after Hemann & Friedman (2013).
;; No disequality constraints.

(provide var var? var=?
         walk unify
         empty-state
         == call/fresh disj conj
         mzero unit mplus bind
         mplus-i bind-i
         (struct-out lazy) peek-weight
         lift-w scale-w mplus-w bind-w)

;; Logic variables are 1-element vectors holding an integer index, so they are
;; distinguishable from other terms (e.g. plain integers) by unification.
(define (var c) (vector c))
(define (var? x) (vector? x))
(define (var=? x1 x2) (= (vector-ref x1 0) (vector-ref x2 0)))

;; A substitution is an association list mapping variables to terms.

(define (assp p l)
  (cond
    [(null? l) #f]
    [(p (car (car l))) (car l)]
    [else (assp p (cdr l))]))

(define (walk u s)
  (let ([pr (and (var? u) (assp (lambda (v) (var=? u v)) s))])
    (if pr (walk (cdr pr) s) u)))

(define (ext-s x v s) (cons (cons x v) s))

(define (unify u v s)
  (let ([u (walk u s)] [v (walk v s)])
    (cond
      [(and (var? u) (var? v) (var=? u v)) s]
      [(var? u) (ext-s u v s)]
      [(var? v) (ext-s v u s)]
      [(and (pair? u) (pair? v))
       (let ([s (unify (car u) (car v) s)])
         (and s (unify (cdr u) (cdr v) s)))]
      [else (and (eqv? u v) s)])))

;; A state is (cons substitution fresh-var-counter).
(define empty-state '(() . 0))

;; Streams: '() is mzero; (cons s/c $) is a mature pair; a procedure of no
;; arguments is an immature stream used for inverse-eta delay.

(define mzero '())
(define (unit s/c) (cons s/c mzero))

(define (mplus $1 $2)
  (cond
    [(null? $1) $2]
    [(procedure? $1) (lambda () (mplus $2 ($1)))]
    [else (cons (car $1) (mplus (cdr $1) $2))]))

(define (bind $ g)
  (cond
    [(null? $) mzero]
    [(procedure? $) (lambda () (bind ($) g))]
    [else (mplus (g (car $)) (bind (cdr $) g))]))

;; Goals are functions from a state to a stream of states.

(define (== u v)
  (lambda (s/c)
    (let ([s (unify u v (car s/c))])
      (if s (unit (cons s (cdr s/c))) mzero))))

(define (call/fresh f)
  (lambda (s/c)
    (let ([c (cdr s/c)])
      ((f (var c)) (cons (car s/c) (+ c 1))))))

(define (disj g1 g2) (lambda (s/c) (mplus (g1 s/c) (g2 s/c))))
(define (conj g1 g2) (lambda (s/c) (bind (g1 s/c) g2)))

;; --- Fair interleaving variants ----------------------------------------
;;
;; Standard `bind` for conj is depth-biased: when stream $ is mature with
;; many states, `(mplus (g (car $)) (bind (cdr $) g))` recursively yields
;; all of (g (car $))'s mature head before the next state from (cdr $).
;; For a recursive relation like `(conj (expr l) (expr r))`, this means
;; the search drills down on (l = first-bank-cell, r = ...) before ever
;; advancing l, producing a strongly depth-biased canonical enumeration
;; order (see bench/deep-bench.rkt for empirical impact).
;;
;; `mplus-i` (interleave) and `bind-i` are the fair variants. After
;; emitting one element from $1, mplus-i suspends the remainder of $1
;; and switches to $2, forcing per-step alternation regardless of
;; whether the underlying streams are mature.
;;
;; References:
;;   Kiselyov, Shan, Friedman, Sabry. "Backtracking, Interleaving, and
;;     Terminating Monad Transformers." ICFP 2005. -- the `interleave`
;;     combinator on MonadPlus from which this is adapted.
;;   Byrd. "Relational Programming in miniKanren: Techniques,
;;     Applications, and Implementations." PhD thesis, Indiana
;;     University, 2009. -- discussion of conjunction fairness and
;;     interleaving alternatives.
;;
;; Note the asymmetry vs `mplus`: the `else` branch wraps the cdr in a
;; thunk so the consumer's next pull will alternate to $2. This costs
;; one thunk allocation per element but produces a fair diagonal
;; enumeration of (l, r) pairs.

(define (mplus-i $1 $2)
  (cond
    [(null? $1) $2]
    [(procedure? $1) (lambda () (mplus-i $2 ($1)))]
    [else (cons (car $1) (lambda () (mplus-i $2 (cdr $1))))]))

(define (bind-i $ g)
  (cond
    [(null? $) mzero]
    [(procedure? $) (lambda () (bind-i ($) g))]
    [else (mplus-i (g (car $)) (bind-i (cdr $) g))]))

;; --- Weighted streams (depth-decayed best-first) ----------------------
;;
;; A weighted stream is one of:
;;   '()                              -- empty
;;   (cons (cons weight state) rest)  -- mature head with explicit weight
;;   (lazy weight thunk)              -- immature, with a known upper bound
;;                                      on weights any forced cell could have
;;
;; Carrying a weight ceiling on immature streams is the critical design
;; choice: `mplus-w` can determine which side could possibly emit a
;; higher-weight cell *without forcing* either side. This avoids the
;; classic "peek requires force" trap that, in a recursive memoized
;; bank, would cause forcing of a memo-thunk while it's still being
;; computed (cycle / infinite recursion).
;;
;; Inspired by best-first proof search in Lean 4's aesop tactic
;; (Limperg & From, "Aesop: White-Box Best-First Proof Search for
;; Lean," CPP 2023), and by admissible heuristics in A*-style search
;; (Hart, Nilsson, Raphael 1968) -- `lazy-weight` plays the role of an
;; admissible upper bound on what the lazy can yield.
;;
;; Weight propagation:
;;   - Zzz-w wraps a goal as `(lazy 1.0 ...)` -- conservative default.
;;   - scale-w multiplies the lazy's ceiling (and each mature cell's
;;     weight) by `factor`.
;;   - bind-w on `(lazy w ...)` produces `(lazy w ...)` -- assumes the
;;     applied goal's output ceiling is <= 1.0 (true for unweighted
;;     goals; weighted goals are conservatively over-estimated).
;;   - mplus-w of two lazies returns a lazy with ceiling = max of the
;;     two ceilings (any cell emitted comes from one of them).
;;   - lift-w of an unweighted procedure stream wraps as `(lazy 1.0 ...)`.

(struct lazy (weight thunk) #:transparent)

(define (peek-weight $)
  (cond
    [(null? $) -inf.0]
    [(lazy? $) (lazy-weight $)]
    [else (caar $)]))

(define (weighted-cell? c)
  (and (pair? c) (pair? (car c)) (number? (caar c))))

;; lift-w : promote any stream to weighted form.
;; - Already weighted (mature with (weight . state) cell, or lazy) -> unchanged.
;; - Unweighted procedure -> wrap as lazy with ceiling 1.0.
;; - Unweighted mature stream -> tag each cell with weight 1.0.
(define (lift-w $)
  (cond
    [(null? $) '()]
    [(lazy? $) $]
    [(procedure? $) (lazy 1.0 (lambda () (lift-w ($))))]
    [(weighted-cell? $) $]
    [else (cons (cons 1.0 (car $)) (lift-w (cdr $)))]))

(define (scale-w factor $)
  (cond
    [(null? $) '()]
    [(lazy? $) (lazy (* factor (lazy-weight $))
                     (lambda () (scale-w factor ((lazy-thunk $)))))]
    [else (cons (cons (* factor (caar $)) (cdar $))
                (scale-w factor (cdr $)))]))

;; mplus-w: emit cells in descending weight order. Uses peek-weight to
;; decide which side wins without forcing the loser. Only forces a lazy
;; when its ceiling indicates it might still produce a winning cell.
;;
;; A merge of many alternatives (one per pending disjunct or bind
;; cell) is kept in a pairing max-heap keyed by peek-weight, rather
;; than in a nest of binary merges: with a binary nest, emitting the
;; next cell costs O(#alternatives), which dominates the run time of
;; best-first search (each pull re-traverses the whole frontier of
;; suspended alternatives). The heap is hidden behind the public
;; stream interface as `lazy-merge`, a subtype of `lazy` that
;; additionally carries the heap, so that nested mplus-w calls meld
;; heaps in O(1) while every other consumer treats the merge as an
;; ordinary immature stream.

(struct lazy-merge lazy (heap) #:transparent)

;; Pairing max-heap of (non-null, non-merge) streams keyed by
;; peek-weight. A heap is #f (empty) or (cons top-stream children),
;; where children is a list of heaps.
(define (heap-meld h1 h2)
  (cond
    [(not h1) h2]
    [(not h2) h1]
    [(>= (peek-weight (car h1)) (peek-weight (car h2)))
     (cons (car h1) (cons h2 (cdr h1)))]
    [else (cons (car h2) (cons h1 (cdr h2)))]))

(define (heap-meld-pairs hs)
  (cond
    [(null? hs) #f]
    [(null? (cdr hs)) (car hs)]
    [else (heap-meld (heap-meld (car hs) (cadr hs))
                     (heap-meld-pairs (cddr hs)))]))

;; heap-add: insert a stream, flattening nested merges by melding
;; their heaps directly.
(define (heap-add h $)
  (cond
    [(null? $) h]
    [(lazy-merge? $) (heap-meld h (lazy-merge-heap $))]
    [else (heap-meld h (cons $ '()))]))

;; heap->stream: expose a heap as a public weighted stream. If the
;; top is mature, emit its head cell and re-insert its tail. If the
;; top is immature, the whole merge is immature with the top's
;; ceiling; forcing it forces exactly one step of the top.
(define (heap->stream h)
  (cond
    [(not h) '()]
    [else
     (let ([top (car h)])
       (cond
         [(lazy? top)
          (lazy-merge (lazy-weight top)
                      (lambda ()
                        (let ([rest (heap-meld-pairs (cdr h))])
                          (heap->stream (heap-add rest ((lazy-thunk top))))))
                      h)]
         [else
          (let ([rest (heap-meld-pairs (cdr h))])
            (cons (car top) (heap->stream (heap-add rest (cdr top)))))]))]))

(define (mplus-w $1 $2)
  (heap->stream (heap-add (heap-add #f $1) $2)))

;; bind-w: for each (w, state) in $, apply g, scale by w. The output
;; ceiling is conservatively (lazy-weight $) -- assumes g's output is
;; bounded by 1.0. -w goals that produce smaller output simply emit
;; cells with lower weight; they still sort correctly because of how
;; scale-w decay propagates.
(define (bind-w $ g)
  (cond
    [(null? $) '()]
    [(lazy? $) (lazy (lazy-weight $)
                     (lambda () (bind-w ((lazy-thunk $)) g)))]
    [else
     (let ([w (caar $)] [s (cdar $)])
       (mplus-w (scale-w w (lift-w (g s)))
                (bind-w (cdr $) g)))]))
\end{minted}

\subsubsection{\texttt{wrappers.rkt}}
\label{sec:appendix-wrappers}

\begin{minted}{racket}
#lang racket/base

;; miniKanren-style wrappers around the microKanren core: variadic
;; conde / fresh, run / run*, plus the stream driver and reifier.
;; Follows section 4 of Hemann & Friedman (2013).

(require (for-syntax racket/base)
         "microkanren.rkt")

(provide Zzz conj+ disj+ conde fresh
         conj-i conj-i+ conde-i fresh-i
         Zzz-w conj-w conj-w+ disj-w disj-w+ conde-w fresh-w
         run run* run-w run*-w
         pull take take-all pull-w take-w take-all-w
         walk*
         current-memo)

;; Memoization cache for defrel/memo, populated per `run` invocation.
;; #f means "no memo session active" -- memoized relations fall back to
;; running their body normally.
(define current-memo (make-parameter #f))

;; Inverse-eta delay: wraps a goal so that its stream is a thunk, which
;; lets disj+/conj+ recur into (potentially nonterminating) goals safely.
(define-syntax Zzz
  (syntax-rules ()
    [(_ g) (lambda (s/c) (lambda () (g s/c)))]))

(define-syntax conj+
  (syntax-rules ()
    [(_ g) (Zzz g)]
    [(_ g0 g ...) (conj (Zzz g0) (conj+ g ...))]))

(define-syntax disj+
  (syntax-rules ()
    [(_ g) (Zzz g)]
    [(_ g0 g ...) (disj (Zzz g0) (disj+ g ...))]))

(define-syntax conde
  (syntax-rules ()
    [(_ (g0 g ...) ...) (disj+ (conj+ g0 g ...) ...)]))

(define-syntax fresh
  (syntax-rules ()
    [(_ () g0 g ...) (conj+ g0 g ...)]
    [(_ (x0 x ...) g0 g ...)
     (call/fresh (lambda (x0) (fresh (x ...) g0 g ...)))]))

;; --- Fair interleaving variants (Kiselyov-Shan-Friedman-Sabry 2005) ---
;;
;; conj-i  : like conj but uses bind-i instead of bind.
;; conj-i+ : variadic, like conj+ but each composition is interleaving.
;; fresh-i : like fresh but combines its body with conj-i+.
;; conde-i : like conde but uses conj-i+ inside each clause.
;;
;; Use these when conj's depth bias hurts -- typically when the body
;; has independent recursive subgoals like `(rel l) (rel r)` and you
;; want fair coverage of (l, r) pairs along antidiagonals rather than
;; a nested-loop enumeration. The cost is one extra thunk allocation
;; per emitted state.

(define (conj-i g1 g2) (lambda (s/c) (bind-i (g1 s/c) g2)))

(define-syntax conj-i+
  (syntax-rules ()
    [(_ g) (Zzz g)]
    [(_ g0 g ...) (conj-i (Zzz g0) (conj-i+ g ...))]))

(define-syntax conde-i
  (syntax-rules ()
    [(_ (g0 g ...) ...) (disj+ (conj-i+ g0 g ...) ...)]))

(define-syntax fresh-i
  (syntax-rules ()
    [(_ () g0 g ...) (conj-i+ g0 g ...)]
    [(_ (x0 x ...) g0 g ...)
     (call/fresh (lambda (x0) (fresh-i (x ...) g0 g ...)))]))

;; --- Weighted (best-first) combinators --------------------------------
;;
;; These operate on weighted streams (see microkanren.rkt). Each goal's
;; output is auto-lifted: == and other unweighted goals produce
;; weight-1 cells when consumed by a -w combinator. Goals returned by
;; defrel/bank-w already produce weighted streams and pass through
;; lift-w unchanged.

;; Zzz-w wraps a goal so its application produces an immature weighted
;; stream with a 1.0 weight ceiling. The ceiling is conservative --
;; a -w goal whose underlying cells are scaled down (e.g. by decay)
;; will report lower actual cell weights once forced, but the lazy's
;; advertised ceiling stays at 1.0 unless we have static info to do
;; better. scale-w can tighten the ceiling at relation boundaries.
(define-syntax Zzz-w
  (syntax-rules ()
    [(_ g) (lambda (s/c) (lazy 1.0 (lambda () (lift-w (g s/c)))))]))

(define (conj-w g1 g2)
  (lambda (s/c) (bind-w (lift-w (g1 s/c)) g2)))

(define (disj-w g1 g2)
  (lambda (s/c) (mplus-w (lift-w (g1 s/c)) (lift-w (g2 s/c)))))

(define-syntax conj-w+
  (syntax-rules ()
    [(_ g) (Zzz-w g)]
    [(_ g0 g ...) (conj-w (Zzz-w g0) (conj-w+ g ...))]))

(define-syntax disj-w+
  (syntax-rules ()
    [(_ g) (Zzz-w g)]
    [(_ g0 g ...) (disj-w (Zzz-w g0) (disj-w+ g ...))]))

(define-syntax conde-w
  (syntax-rules ()
    [(_ (g0 g ...) ...) (disj-w+ (conj-w+ g0 g ...) ...)]))

(define-syntax fresh-w
  (syntax-rules ()
    [(_ () g0 g ...) (conj-w+ g0 g ...)]
    [(_ (x0 x ...) g0 g ...)
     (call/fresh (lambda (x0) (fresh-w (x ...) g0 g ...)))]))

;; Weighted reify: extract state from weighted cell before reifying.
(define (reify-1st/w cell)
  (let* ([s/c (cdr cell)]
         [v (walk* (var 0) (car s/c))])
    (walk* v (reify-s v '()))))

;; pull/take for weighted streams: lazy structs (not procedures) signal
;; immature; force via lazy-thunk.
(define (pull-w $)
  (cond
    [(lazy? $) (pull-w ((lazy-thunk $)))]
    [else $]))

(define (take-w n $)
  (if (zero? n) '()
      (let ([$ (pull-w $)])
        (if (null? $) '()
            (cons (car $) (take-w (- n 1) (cdr $)))))))

(define (take-all-w $)
  (let ([$ (pull-w $)])
    (if (null? $) '()
        (cons (car $) (take-all-w (cdr $))))))

(define-syntax run-w
  (syntax-rules ()
    [(_ n (q) g0 g ...)
     (parameterize ([current-memo (make-hash)])
       (map reify-1st/w (take-w n ((fresh-w (q) g0 g ...) empty-state))))]
    [(_ n (x0 x ...) g0 g ...)
     (run-w n (q) (fresh-w (x0 x ...) (== q (list x0 x ...)) g0 g ...))]))

(define-syntax run*-w
  (syntax-rules ()
    [(_ (q) g0 g ...)
     (parameterize ([current-memo (make-hash)])
       (map reify-1st/w (take-all-w ((fresh-w (q) g0 g ...) empty-state))))]
    [(_ (x0 x ...) g0 g ...)
     (run*-w (q) (fresh-w (x0 x ...) (== q (list x0 x ...)) g0 g ...))]))

;; --- stream driver ---

(define (pull $) (if (procedure? $) (pull ($)) $))

(define (take-all $)
  (let ([$ (pull $)])
    (if (null? $) '()
        (cons (car $) (take-all (cdr $))))))

(define (take n $)
  (if (zero? n) '()
      (let ([$ (pull $)])
        (if (null? $) '()
            (cons (car $) (take (- n 1) (cdr $)))))))

;; --- reification ---

(define (walk* v s)
  (let ([v (walk v s)])
    (cond
      [(var? v) v]
      [(pair? v) (cons (walk* (car v) s) (walk* (cdr v) s))]
      [else v])))

(define (reify-name n)
  (string->symbol (string-append "_." (number->string n))))

(define (reify-s v s)
  (let ([v (walk v s)])
    (cond
      [(var? v) (cons (cons v (reify-name (length s))) s)]
      [(pair? v) (reify-s (cdr v) (reify-s (car v) s))]
      [else s])))

;; The first fresh variable introduced from empty-state has index 0;
;; the run macro arranges for that variable to hold the query value.
(define (reify-1st s/c)
  (let ([v (walk* (var 0) (car s/c))])
    (walk* v (reify-s v '()))))

;; --- run / run* ---
;;
;; Single-var form returns reified values directly.
;; Multi-var form bundles the query vars into a list under a hygienic q.

(define-syntax run
  (syntax-rules ()
    [(_ n (q) g0 g ...)
     (parameterize ([current-memo (make-hash)])
       (map reify-1st (take n ((fresh (q) g0 g ...) empty-state))))]
    [(_ n (x0 x ...) g0 g ...)
     (run n (q) (fresh (x0 x ...) (== q (list x0 x ...)) g0 g ...))]))

(define-syntax run*
  (syntax-rules ()
    [(_ (q) g0 g ...)
     (parameterize ([current-memo (make-hash)])
       (map reify-1st (take-all ((fresh (q) g0 g ...) empty-state))))]
    [(_ (x0 x ...) g0 g ...)
     (run* (q) (fresh (x0 x ...) (== q (list x0 x ...)) g0 g ...))]))
\end{minted}

\subsubsection{\texttt{prune.rkt}}
\label{sec:appendix-prune}

\begin{minted}{racket}
#lang racket/base

;; Pruning combinator.
;;
;; (prune key g) wraps a goal g and filters its answer stream so that at
;; most one state is emitted per distinct value of (key s/c). The key
;; function is supplied per call, so the equivalence used to prune is
;; chosen locally -- e.g. "behavior on the input examples" for PBE
;; synthesis, or "shape modulo alpha-renaming" elsewhere.
;;
;; The dedup table is local to one prune call and shared across the
;; stream's lazy thunks (the closure captures it), so dedup state
;; survives the inverse-eta delay used by mplus/bind.
;;
;; This sketch uses equal?-keyed hashing; an efficient version would
;; hash-cons the key values produced by the user's semantic function.

(require "microkanren.rkt"
         "wrappers.rkt")

(provide prune skip-prune
         ground? ground-key when-ground
         prune-w)

;; Sentinel a key function may return to mean "don't dedup yet" --
;; typically because the relevant variables are still fresh.
(define skip-prune 'skip-prune)

;; prune : (state -> any) goal -> goal
(define (prune key g)
  (lambda (s/c)
    (prune-stream key (make-hash) (g s/c))))

(define (prune-stream key seen $)
  (cond
    [(null? $) '()]
    [(procedure? $) (lambda () (prune-stream key seen ($)))]
    [else
     (let* ([s/c (car $)]
            [k (key s/c)])
       (cond
         [(eq? k skip-prune)
          (cons s/c (prune-stream key seen (cdr $)))]
         [(hash-has-key? seen k)
          (prune-stream key seen (cdr $))]
         [else
          (hash-set! seen k #t)
          (cons s/c (prune-stream key seen (cdr $)))]))]))

;; --- key-building helpers ---

;; A term is ground if it contains no logic variables.
(define (ground? t)
  (cond
    [(var? t) #f]
    [(pair? t) (and (ground? (car t)) (ground? (cdr t)))]
    [else #t]))

;; ground-key : variable (term -> any) -> (state -> any)
;;
;; Lifts a host function on terms into a prune key on states: walks `v`
;; in the current substitution and, if the result is ground, applies `f`
;; to it. Returns `skip-prune` when `v` hasn't ground out yet, so those
;; states pass through `prune` unfiltered.
;;
;; Typical use in PBE synthesis:
;;   (prune (ground-key e (lambda (t) (map (eval-on t) inputs)))
;;          (expr e depth))
(define (ground-key v f)
  (lambda (s/c)
    (let ([t (walk* v (car s/c))])
      (if (ground? t)
          (f t)
          skip-prune))))

;; when-ground : variable (term -> boolean) -> goal
;;
;; A goal that succeeds when `v` walks to a ground term satisfying
;; `pred`, and fails otherwise (including when `v` is still non-ground).
;; Useful as the final "accept this candidate" stage of a synthesis run.
(define (when-ground v pred)
  (lambda (s/c)
    (let ([t (walk* v (car s/c))])
      (if (and (ground? t) (pred t))
          (unit s/c)
          '()))))

;; --- prune-w : prune for weighted streams ------------------------------

(define (prune-w key g)
  (lambda (s/c)
    (prune-stream-w key (make-hash) (lift-w (g s/c)))))

(define (prune-stream-w key seen $)
  (cond
    [(null? $) '()]
    [(lazy? $) (lazy (lazy-weight $)
                     (lambda () (prune-stream-w key seen ((lazy-thunk $)))))]
    [else
     (let* ([cell (car $)] [s (cdr cell)] [k (key s)])
       (cond
         [(eq? k skip-prune)
          (cons cell (prune-stream-w key seen (cdr $)))]
         [(hash-has-key? seen k)
          (prune-stream-w key seen (cdr $))]
         [else
          (hash-set! seen k #t)
          (cons cell (prune-stream-w key seen (cdr $)))]))]))
\end{minted}

\subsubsection{\texttt{memo.rkt}}
\label{sec:appendix-memo}

\begin{minted}{racket}
#lang racket/base

;; defrel/memo: define a pure relation whose canonical answer stream is
;; computed once per `run` and replayed against each caller's arguments.
;;
;; The body of a memoized relation is run with canonical input vars
;; (var 0), (var 1), ..., (var N-1) (N = arity), starting counter N.
;; The resulting stream of states (lazy) is stored in the per-run cache.
;; Each subsequent call replays this canonical stream against the
;; caller's args:
;;   - canonical input vars are renamed to the caller's actual args;
;;   - canonical internal fresh vars (idx >= N) are shifted by the
;;     caller's counter to allocate fresh vars in the caller's namespace;
;;   - the renamed bindings are unified into the caller's substitution.
;;
;; Recursive calls inside the body Zzz-suspend, so by the time they fire
;; the cache entry is in place. The lazy thunks of the canonical stream
;; are memoized (`memo-thunk`) so forcing happens at most once per cell
;; across all replays.

(require "microkanren.rkt"
         "wrappers.rkt"
         "prune.rkt")

(provide defrel/memo defrel/bank defrel/bank-w with-memo-session)

(define-syntax-rule (with-memo-session body ...)
  (parameterize ([current-memo (make-hash)]) body ...))

;; --- thunk + stream memoization ---

(define (memo-thunk f)
  (let ([result #f] [done? #f])
    (lambda ()
      (unless done?
        (set! result (f))
        (set! done? #t))
      result)))

(define (memo-stream s)
  (cond
    [(null? s) '()]
    ;; Weighted: preserve the lazy's weight ceiling, but memoize the
    ;; forcing of its thunk. Each lazy gets its own memo-thunk; multiple
    ;; replays sharing the same canonical bank all hit the cache.
    [(lazy? s) (lazy (lazy-weight s)
                     (memo-thunk (lambda () (memo-stream ((lazy-thunk s))))))]
    [(procedure? s) (memo-thunk (lambda () (memo-stream (s))))]
    [else (cons (car s) (memo-stream (cdr s)))]))


;; --- replay ---

(define (replay-state canonical-state caller-state args-vec num-args)
  (define canonical-subst (car canonical-state))
  (define canonical-counter (cdr canonical-state))
  (define caller-subst (car caller-state))
  (define caller-counter (cdr caller-state))
  (define shift-internal (- caller-counter num-args))
  (define new-counter (+ caller-counter (- canonical-counter num-args)))
  (define (rename t)
    (cond
      [(var? t)
       (let ([idx (vector-ref t 0)])
         (cond
           [(< idx num-args) (vector-ref args-vec idx)]
           [else (var (+ idx shift-internal))]))]
      [(pair? t) (cons (rename (car t)) (rename (cdr t)))]
      [else t]))
  (let loop ([i 0] [s caller-subst])
    (cond
      [(= i num-args) (cons s new-counter)]
      [else
       (let* ([canonical-value (walk* (var i) canonical-subst)]
              [renamed (rename canonical-value)]
              [caller-arg (vector-ref args-vec i)]
              [s* (unify caller-arg renamed s)])
         (if s* (loop (+ i 1) s*) #f))])))

(define (replay-stream canonical caller-state args-vec num-args)
  (cond
    [(null? canonical) '()]
    [(procedure? canonical)
     ;; Collapse a chain of immature canonical thunks into a single
     ;; replay-thunk force -- when the consumer asks for the next state,
     ;; we keep forcing until we get a concrete cons (or null), instead
     ;; of producing a new replay-thunk for each layer of canonical
     ;; laziness. This is what `pull` does on the consumer side; doing
     ;; it here cuts the thunk-force count from ~330k to ~3k for the
     ;; PBE bench and brings defrel/bank to within 1.5x of the
     ;; depth-bounded baseline.
     (lambda ()
       (let loop ([c (canonical)])
         (cond
           [(null? c) '()]
           [(procedure? c) (loop (c))]
           [else
            (let ([new (replay-state (car c) caller-state args-vec num-args)])
              (cond
                [new (cons new (replay-stream (cdr c) caller-state args-vec num-args))]
                [else (replay-stream (cdr c) caller-state args-vec num-args)]))])))]
    [else
     (let ([new (replay-state (car canonical) caller-state args-vec num-args)])
       (cond
         [new
          (cons new
                (replay-stream (cdr canonical) caller-state args-vec num-args))]
         [else
          (replay-stream (cdr canonical) caller-state args-vec num-args)]))]))

;; --- init + memo-call ---

(define (init-cell cache rel-tag body-fn num-args)
  (define canonical-vars (for/list ([i (in-range num-args)]) (var i)))
  (define canonical-state (cons '() num-args))
  (define b (box #f))
  (hash-set! cache rel-tag b)
  (define s ((apply body-fn canonical-vars) canonical-state))
  (set-box! b (memo-stream s))
  b)

(define (make-memo-rel num-args body-fn)
  (define rel-tag (box #f))
  (lambda actual-args
    (let ([args-vec (list->vector actual-args)])
      (lambda (s/c)
        (let ([cache (current-memo)])
          (cond
            [cache
             (let ([b (or (hash-ref cache rel-tag #f)
                          (init-cell cache rel-tag body-fn num-args))])
               (replay-stream (unbox b) s/c args-vec num-args))]
            [else
             ((apply body-fn actual-args) s/c)]))))))

(define-syntax defrel/memo
  (syntax-rules ()
    [(_ (name x ...) body ...)
     (define name
       (make-memo-rel (length '(x ...))
                      (lambda (x ...) (conj+ body ...))))]))

;; defrel/bank: like defrel/memo but additionally prunes the canonical
;; stream by the supplied key. Prune runs once at canonical-stream
;; construction time; replays already see only one representative per
;; key. Use shape:
;;
;;   (defrel/bank (rel x ...) #:prune key-expr body ...)
;;
;; key-expr is evaluated in the scope of the relation parameters, so it
;; can reference them (e.g. (ground-key x (lambda (t) ...))). The
;; canonical run binds the parameters to (var 0), (var 1), ....

(define (init-bank-cell cache rel-tag body-fn key-fn num-args)
  (define canonical-vars (for/list ([i (in-range num-args)]) (var i)))
  (define canonical-state (cons '() num-args))
  (define b (box #f))
  (hash-set! cache rel-tag b)
  (define key (apply key-fn canonical-vars))
  (define body-goal (apply body-fn canonical-vars))
  (define s ((prune key body-goal) canonical-state))
  (set-box! b (memo-stream s))
  b)

(define (make-bank-rel num-args body-fn key-fn)
  (define rel-tag (box #f))
  (lambda actual-args
    (let ([args-vec (list->vector actual-args)])
      (lambda (s/c)
        (let ([cache (current-memo)])
          (cond
            [cache
             (let ([b (or (hash-ref cache rel-tag #f)
                          (init-bank-cell cache rel-tag body-fn key-fn num-args))])
               (replay-stream (unbox b) s/c args-vec num-args))]
            [else
             ;; No memo session: fall back to pruning per-call.
             (let ([k (apply key-fn actual-args)]
                   [g (apply body-fn actual-args)])
               ((prune k g) s/c))]))))))

(define-syntax defrel/bank
  (syntax-rules ()
    [(_ (name x ...) #:prune key-expr body ...)
     (define name
       (make-bank-rel (length '(x ...))
                      (lambda (x ...) (conj+ body ...))
                      (lambda (x ...) key-expr)))]))

;; --- defrel/bank-w : weighted bank with depth decay --------------------
;;
;; Same as defrel/bank but uses weighted streams and applies a decay
;; factor to each call. Recursive uses of the relation get weight
;; scaled by `decay` (default 0.5). With sorted-merge mplus-w, this
;; produces depth-ordered enumeration: shallow representatives are
;; emitted before deeper ones.
;;
;;   (defrel/bank-w (rel x ...) #:prune key-expr #:decay d body ...)
;;
;; The body should use the *-w combinators (conde-w, fresh-w, conj-w+)
;; so its internal scheduling participates in the weighted ordering.
;; Plain == is auto-lifted to weight 1.
;;
;; Replay-stream-w is the weighted analogue of replay-stream: each
;; canonical cell carries a weight that gets passed through to the
;; caller. The outer `(scale-w decay ...)` wrapper adds one factor of
;; decay per invocation, so a depth-K canonical cell delivered to the
;; caller has weight roughly decay^(2K+1).

(define (replay-stream-w canonical caller-state args-vec num-args)
  (cond
    [(null? canonical) '()]
    [(lazy? canonical)
     ;; Preserve the canonical lazy's weight ceiling -- replay doesn't
     ;; change weights, only renames vars. Crucially, we do NOT force
     ;; the canonical thunk here; the lazy is returned as-is, and the
     ;; consumer forces it only when its weight indicates it might win.
     (lazy (lazy-weight canonical)
           (lambda ()
             (replay-stream-w ((lazy-thunk canonical))
                              caller-state args-vec num-args)))]
    [else
     (let* ([cell (car canonical)] [w (car cell)] [cs (cdr cell)])
       (let ([new (replay-state cs caller-state args-vec num-args)])
         (cond
           [new (cons (cons w new)
                      (replay-stream-w (cdr canonical) caller-state args-vec num-args))]
           [else (replay-stream-w (cdr canonical) caller-state args-vec num-args)])))]))

(define (init-bank-cell-w cache rel-tag body-fn key-fn num-args)
  (define canonical-vars (for/list ([i (in-range num-args)]) (var i)))
  (define canonical-state (cons '() num-args))
  (define b (box #f))
  (hash-set! cache rel-tag b)
  (define key (apply key-fn canonical-vars))
  (define body-goal (apply body-fn canonical-vars))
  (define s ((prune-w key body-goal) canonical-state))
  (set-box! b (memo-stream s))
  b)

(define (make-bank-rel-w num-args body-fn key-fn decay)
  (define rel-tag (box #f))
  (lambda actual-args
    (let ([args-vec (list->vector actual-args)])
      (lambda (s/c)
        (let ([cache (current-memo)])
          (cond
            [cache
             (let ([b (or (hash-ref cache rel-tag #f)
                          (init-bank-cell-w cache rel-tag body-fn key-fn num-args))])
               (scale-w decay
                 (replay-stream-w (unbox b) s/c args-vec num-args)))]
            [else
             (scale-w decay
               ((prune-w (apply key-fn actual-args) (apply body-fn actual-args)) s/c))]))))))

(define-syntax defrel/bank-w
  (syntax-rules ()
    [(_ (name x ...) #:prune key-expr #:decay decay body ...)
     (define name
       (make-bank-rel-w (length '(x ...))
                        (lambda (x ...) (conj-w+ body ...))
                        (lambda (x ...) key-expr)
                        decay))]
    [(_ (name x ...) #:prune key-expr body ...)
     ;; default decay
     (defrel/bank-w (name x ...) #:prune key-expr #:decay 0.5 body ...)]))
\end{minted}

\subsubsection{\texttt{bench/bank.rkt}}
\label{sec:appendix-bank}

\begin{minted}{racket}
#lang racket/base

;; Bottom-up observational-equivalence bank for PBE synthesis.
;;
;; build-bank/depth grows a set of programs layer by layer, keeping one
;; representative per observable behavior. Composes in host code; the
;; produced bank is then enumerated relationally by `membero`.
;;
;; This is the standard PBE bank construction. It preserves completeness
;; under conj/bind (unlike a shared dedup table) because each entry in
;; the bank is a concrete syntactic rep that can be replayed any number
;; of times for any caller.

(require "../microkanren.rkt")

(provide build-bank/depth membero)

;; build-bank/depth :
;;   (listof prog -> listof prog)   ; grammar-step: compose bank entries
;;   (listof prog)                  ; terminals: leaf programs
;;   (prog -> any)                  ; behavior: observable key
;;   natural                        ; d: number of composition layers
;;   -> (listof prog)
(define (build-bank/depth grammar-step terminals behavior d)
  (define seen (make-hash))
  (define bank '())
  (define (try p)
    (define b (behavior p))
    (unless (hash-has-key? seen b)
      (hash-set! seen b #t)
      (set! bank (cons p bank))))
  (for ([t (in-list terminals)]) (try t))
  (for ([_ (in-range d)])
    (define snapshot bank)  ; freeze the bank for this layer's compositions
    (for ([p (in-list (grammar-step snapshot))]) (try p)))
  (reverse bank))

;; membero : term (listof term) -> goal
;;
;; A goal that succeeds with v unified to each element of ls in turn.
;; Lazy: produces one state per pull, so `run 1` stops at the first
;; match without materializing the whole list of states.
(define (membero v ls)
  (lambda (s/c)
    (let loop ([ls ls])
      (cond
        [(null? ls) '()]
        [else
         (let ([result ((== v (car ls)) s/c)])
           (cond
             [(null? result) (loop (cdr ls))]
             [else (cons (car result) (lambda () (loop (cdr ls))))]))]))))
\end{minted}

\subsubsection{\texttt{main.rkt}}
\label{sec:appendix-main}

\begin{minted}{racket}
#lang racket/base

;; prune-kanren: miniKanren with built-in pruning for PBE program synthesis.
;; Re-exports the microKanren core, the miniKanren-style surface forms,
;; the prune combinator, and the memoized relation forms.

(require "microkanren.rkt"
         "wrappers.rkt"
         "prune.rkt"
         "memo.rkt")
(provide (all-from-out "microkanren.rkt")
         (all-from-out "wrappers.rkt")
         (all-from-out "prune.rkt")
         (all-from-out "memo.rkt"))
\end{minted}

\subsubsection{\texttt{info.rkt}}
\label{sec:appendix-info}

\begin{minted}{racket}
#lang info

(define collection "prune-kanren")
(define pkg-desc
  "A miniKanren implementation with a built-in pruning mechanism, suitable for PBE program synthesis.")
(define version "0.0")
(define pkg-authors '(nikolai-kudasov))
(define license 'MIT)

(define deps '("base"))
(define build-deps '("racket-doc"
                     "rackunit-lib"
                     "scribble-lib"))
\end{minted}

\subsubsection{\texttt{examples/shared-table-witness.rkt}}
\label{sec:appendix-witness-listing}

\begin{minted}{racket}
#lang racket/base

;; Experimental: shared dedup table across nested prunes via a parameter.
;; The outermost prune call creates the table; recursive prune calls
;; inherit it via current-prune-table. Tests whether sharing speeds up
;; the depth-less search AND preserves the synthesis answer.
;;
;; RESULT: sharing breaks completeness. Running this file hangs on the
;; easy target (`x*x`) before reaching the hard one. The reason: the
;; outer prune emits the leaf `(== e 'x)` first, adding behavior (2 3 4)
;; to the shared table. When the recursive branch then runs `(expr l)`
;; and `(expr r)` to try to form `(times x x)`, both children would be
;; `x` with behavior (2 3 4) -- which is now blocked by the table.
;; (times x x) is unreachable, and the search never finds another ground
;; term with behavior (4 9 16) so it loops forever pulling at the
;; depth-less stream.
;;
;; More generally: simple sharing breaks any synthesis target whose
;; sub-expressions share a behavior with anything already emitted.
;; This benchmark is kept as a witness to that failure. It is NOT meant
;; to be run unattended; the second run will hang.

(require "../main.rkt")

;; --- shared-table prune ---

(define current-prune-table (make-parameter #f))

(define (prune/shared key g)
  (let ([table (or (current-prune-table) (make-hash))])
    (lambda (s/c)
      (parameterize ([current-prune-table table])
        (prune-stream/shared key table (g s/c))))))

(define (prune-stream/shared key seen $)
  (cond
    [(null? $) '()]
    [(procedure? $)
     (lambda ()
       (parameterize ([current-prune-table seen])
         (prune-stream/shared key seen ($))))]
    [else
     (let* ([s/c (car $)]
            [k (key s/c)])
       (cond
         [(eq? k skip-prune)
          (cons s/c (prune-stream/shared key seen (cdr $)))]
         [(hash-has-key? seen k)
          (prune-stream/shared key seen (cdr $))]
         [else
          (hash-set! seen k #t)
          (cons s/c (prune-stream/shared key seen (cdr $)))]))]))

;; --- problem ---

(define inputs '(2 3 4))
(define easy-io '((2 . 4) (3 . 9) (4 . 16)))
(define hard-io '((2 . 10) (3 . 17) (4 . 26)))

(define (interp e x)
  (cond
    [(eq? e 'x) x]
    [(number? e) e]
    [(and (pair? e) (eq? (car e) 'plus))
     (+ (interp (cadr e) x) (interp (caddr e) x))]
    [(and (pair? e) (eq? (car e) 'times))
     (* (interp (cadr e) x) (interp (caddr e) x))]
    [else (error 'interp "bad expression: ~v" e)]))

(define (behavior-key e)
  (ground-key e (lambda (t) (map (lambda (x) (interp t x)) inputs))))

(define (matches e io-pairs)
  (when-ground e
    (lambda (t)
      (andmap (lambda (io) (equal? (interp t (car io)) (cdr io)))
              io-pairs))))

;; --- expr with each prune flavor ---

(define (expr-orig e)
  (prune (behavior-key e)
    (conde
      [(== e 'x)] [(== e 0)] [(== e 1)]
      [(fresh (l r)
         (conde
           [(== e `(plus  ,l ,r))]
           [(== e `(times ,l ,r))])
         (expr-orig l) (expr-orig r))])))

(define (expr-shared e)
  (prune/shared (behavior-key e)
    (conde
      [(== e 'x)] [(== e 0)] [(== e 1)]
      [(fresh (l r)
         (conde
           [(== e `(plus  ,l ,r))]
           [(== e `(times ,l ,r))])
         (expr-shared l) (expr-shared r))])))

(define (time-it label thunk)
  (collect-garbage)
  (define start (current-inexact-milliseconds))
  (define result (thunk))
  (define elapsed (- (current-inexact-milliseconds) start))
  (printf "  ~a: result=~v  (~a ms)~n"
          label result (real->decimal-string elapsed 1))
  ;; Return void: a non-void value at module level would print twice.
  (void))

(module+ main
  (printf "=== easy target: x*x  (expected: (times x x)) ===~n")
  (time-it "expr-orig    " (lambda () (run 1 (e) (expr-orig e) (matches e easy-io))))
  (time-it "expr-shared  " (lambda () (run 1 (e) (expr-shared e) (matches e easy-io))))

  (printf "~n=== hard target: (1+x)^2+1 ===~n")
  (time-it "expr-orig    " (lambda () (run 1 (e) (expr-orig e) (matches e hard-io))))
  (time-it "expr-shared  " (lambda () (run 1 (e) (expr-shared e) (matches e hard-io)))))
\end{minted}

\subsubsection{\texttt{bench/shallow-bench.rkt}}
\label{sec:appendix-shallow-bench}

\begin{minted}{racket}
#lang racket/base

;; Synthesis benchmark suite: defrel/bank vs depth-bounded prune.
;;
;; Compares the clean depth-less defrel/bank style against the more
;; verbose expr-bounded idiom across multiple PBE targets, in two
;; domains:
;;
;;   1. Arithmetic -- polynomials over a single integer input x.
;;   2. String manipulation -- concatenation grammar over a single
;;      string input, à la Polikarpova's "Big Ideas in Program
;;      Synthesis" examples (greeting formatters, echo patterns, etc).
;;
;; For each target, bounded is run at the smallest depth that admits a
;; solution (best case for bounded). defrel/bank has no depth knob.

(require "../main.rkt"
         "bank.rkt")

;; --- per-target timeout via thread + custodian (copied from deep-bench) ---

(define (with-time-limit ms thunk on-timeout)
  (define result-ch (make-channel))
  (define cust (make-custodian))
  (parameterize ([current-custodian cust])
    (thread
      (lambda ()
        (with-handlers ([exn:fail? (lambda (e) (channel-put result-ch (cons 'error e)))])
          (channel-put result-ch (cons 'ok (thunk)))))))
  (sync
    (handle-evt result-ch
      (lambda (v)
        (case (car v)
          [(ok) (cdr v)]
          [(error) (raise (cdr v))])))
    (handle-evt (alarm-evt (+ (current-inexact-milliseconds) ms))
      (lambda (_)
        (custodian-shutdown-all cust)
        (on-timeout)))))

(define (timed-run budget-ms thunk)
  (collect-garbage)
  (define start (current-inexact-milliseconds))
  (define result
    (with-time-limit budget-ms thunk (lambda () 'TIMEOUT)))
  (define elapsed (- (current-inexact-milliseconds) start))
  (values result elapsed))

(define (time-it iters thunk)
  (collect-garbage)
  (define start (current-inexact-milliseconds))
  (define result #f)
  (for ([i (in-range iters)])
    (set! result (thunk)))
  (define elapsed (- (current-inexact-milliseconds) start))
  (values result (/ elapsed iters)))

(define (pad-right n s)
  (if (>= (string-length s) n)
      s
      (string-append s (make-string (- n (string-length s)) #\space))))

(define (run-target target target-bank target-bounded)
  (define name (car target))
  (define io-pairs (cadr target))
  (define depth (caddr target))
  (define iters (cadddr target))
  (define-values (bank-result bank-ms) (time-it iters target-bank))
  (define-values (bounded-result bounded-ms) (time-it iters target-bounded))
  (printf "  ~a  depth=~a   bounded ~ams   bank ~ams   ratio ~ax~n"
          (pad-right 24 name)
          depth
          (pad-right 7 (real->decimal-string bounded-ms 3))
          (pad-right 7 (real->decimal-string bank-ms 3))
          (real->decimal-string (/ bank-ms bounded-ms) 2))
  (unless (and (pair? bank-result) (pair? bounded-result))
    (printf "    !!! one of the searches returned no answer~n"))
  (printf "    bounded found: ~v~n" (and (pair? bounded-result) (car bounded-result)))
  (printf "    bank    found: ~v~n" (and (pair? bank-result) (car bank-result))))

;; ============================================================
;; Domain 1: Arithmetic
;; ============================================================

(define arith-inputs '(2 3 4))

(define (arith-interp e x)
  (cond
    [(eq? e 'x) x]
    [(number? e) e]
    [(eq? (car e) 'plus)  (+ (arith-interp (cadr e) x) (arith-interp (caddr e) x))]
    [(eq? (car e) 'times) (* (arith-interp (cadr e) x) (arith-interp (caddr e) x))]))

(define (arith-key e)
  (ground-key e (lambda (t) (map (lambda (x) (arith-interp t x)) arith-inputs))))

(define (arith-matches e io-pairs)
  (when-ground e (lambda (t)
    (andmap (lambda (io) (equal? (arith-interp t (car io)) (cdr io))) io-pairs))))

(defrel/bank (arith-bank e)
  #:prune (arith-key e)
  (conde
    [(== e 'x)] [(== e 0)] [(== e 1)]
    [(fresh (l r) (conde [(== e `(plus ,l ,r))] [(== e `(times ,l ,r))])
       (arith-bank l) (arith-bank r))]))

(define (arith-bounded e depth)
  (prune (arith-key e)
    (cond
      [(zero? depth) (conde [(== e 'x)] [(== e 0)] [(== e 1)])]
      [else
       (conde
         [(== e 'x)] [(== e 0)] [(== e 1)]
         [(fresh (l r) (conde [(== e `(plus ,l ,r))] [(== e `(times ,l ,r))])
            (arith-bounded l (- depth 1)) (arith-bounded r (- depth 1)))])])))

;; (name io-pairs minimum-depth iterations)
(define arith-targets
  `(("identity: x"
     ((2 . 2) (3 . 3) (4 . 4))                       0 1000)
    ("square: x*x"
     ((2 . 4) (3 . 9) (4 . 16))                      1 500)
    ("x^2 + x"
     ((2 . 6) (3 . 12) (4 . 20))                     2 200)
    ("x^2 + 2x + 1 = (x+1)^2"
     ((2 . 9) (3 . 16) (4 . 25))                     2 200)
    ("(1+x)^2 + 1"
     ((2 . 10) (3 . 17) (4 . 26))                    3 50)
    ("x^3"
     ((2 . 8) (3 . 27) (4 . 64))                     2 100)
    ("x^3 + 1"
     ((2 . 9) (3 . 28) (4 . 65))                     3 50)))

(define (arith-bench)
  (printf "~n============================================================~n")
  (printf "Arithmetic synthesis  (inputs: ~v)~n" arith-inputs)
  (printf "Grammar: e ::= x | 0 | 1 | (plus e e) | (times e e)~n")
  (printf "============================================================~n")
  (for ([target (in-list arith-targets)])
    (define io-pairs (cadr target))
    (define depth (caddr target))
    (run-target target
      (lambda () (run 1 (e) (arith-bank e) (arith-matches e io-pairs)))
      (lambda () (run 1 (e) (arith-bounded e depth) (arith-matches e io-pairs))))))

;; ============================================================
;; Domain 2: String manipulation (Polikarpova-style)
;; ============================================================

(define str-inputs '("world" "Alice" "Bob"))

(define (str-interp e s)
  (cond
    [(eq? e 'in) s]
    [(string? e) e]
    [(eq? (car e) 'concat)
     (string-append (str-interp (cadr e) s) (str-interp (caddr e) s))]))

(define (str-key e)
  (ground-key e (lambda (t) (map (lambda (s) (str-interp t s)) str-inputs))))

(define (str-matches e io-pairs)
  (when-ground e (lambda (t)
    (andmap (lambda (io) (equal? (str-interp t (car io)) (cdr io))) io-pairs))))

;; String literals available as terminals. In a real PBE system these
;; would be mined from the I/O examples (Polikarpova's approach uses
;; substrings of the outputs that aren't substrings of the inputs).
(define str-literals '("" " " "!" "?" ", " "Hello, " "Hi, " "Dear "))

(defrel/bank (str-bank e)
  #:prune (str-key e)
  (conde
    [(== e 'in)]
    [(membero e str-literals)]
    [(fresh (l r) (== e `(concat ,l ,r))
       (str-bank l) (str-bank r))]))

;; Weighted variant for best-first enumeration. Literals must be
;; spelled out as conde-w branches so that each contributes a uniform
;; ceiling weight; recursive concat branches get scale-w'd by the
;; default decay.
(defrel/bank-w (str-bank-w e)
  #:prune (str-key e)
  #:decay 0.5
  (conde-w
    [(== e 'in)]
    [(== e "")]        [(== e " ")]    [(== e "!")]       [(== e "?")]
    [(== e ", ")]      [(== e "Hello, ")]
    [(== e "Hi, ")]    [(== e "Dear ")]
    [(fresh-w (l r) (== e `(concat ,l ,r))
       (str-bank-w l) (str-bank-w r))]))

(define (str-bounded e depth)
  (prune (str-key e)
    (cond
      [(zero? depth)
       (conde [(== e 'in)] [(membero e str-literals)])]
      [else
       (conde
         [(== e 'in)]
         [(membero e str-literals)]
         [(fresh (l r) (== e `(concat ,l ,r))
            (str-bounded l (- depth 1)) (str-bounded r (- depth 1)))])])))

;; Host-language bottom-up bank for strings.
(define str-terminals (cons 'in str-literals))

(define (str-grammar-step bank)
  (for*/list ([l (in-list bank)]
              [r (in-list bank)])
    (list 'concat l r)))

(define (str-behavior p) (map (lambda (s) (str-interp p s)) str-inputs))

(define (str-host-bank-result depth io-pairs)
  (define bank (build-bank/depth str-grammar-step str-terminals
                                 str-behavior depth))
  (run 1 (e) (membero e bank) (str-matches e io-pairs)))

(define str-targets
  `(("greeting: 'Hello, X!'"
     (("world" . "Hello, world!")
      ("Alice" . "Hello, Alice!")
      ("Bob"   . "Hello, Bob!"))
     2 100)
    ("informal greeting: 'Hi, X?'"
     (("world" . "Hi, world?")
      ("Alice" . "Hi, Alice?")
      ("Bob"   . "Hi, Bob?"))
     2 100)
    ("echo-with-comma: 'X, X'"
     (("world" . "world, world")
      ("Alice" . "Alice, Alice")
      ("Bob"   . "Bob, Bob"))
     2 200)
    ("greeting+shout: 'Hello, X!!'"
     (("world" . "Hello, world!!")
      ("Alice" . "Hello, Alice!!")
      ("Bob"   . "Hello, Bob!!"))
     3 10)))

(define (fmt-result v)
  (cond
    [(eq? v 'TIMEOUT) "TIMEOUT"]
    [(pair? v) (format "~v" (car v))]
    [(null? v) "no-answer"]
    [else (format "~v" v)]))

;; Per-target budget for the bank-w engine on string targets. bank-w
;; can be very slow on wide behavior spaces, so we wrap it in a
;; timeout rather than relying on the iteration count to bound it.
(define str-bank-w-budget-ms 30000)

(define (str-bench)
  (printf "~n============================================================~n")
  (printf "String synthesis  (inputs: ~v)~n" str-inputs)
  (printf "Grammar: e ::= in | <literal> | (concat e e)~n")
  (printf "Literals: ~v~n" str-literals)
  (printf "============================================================~n")
  (for ([target (in-list str-targets)])
    (define name (car target))
    (define io-pairs (cadr target))
    (define depth (caddr target))
    (define iters (cadddr target))
    (printf "  ~a   depth=~a  (iters=~a)~n"
            (pad-right 30 name) depth iters)
    (flush-output)

    (define-values (bounded-r bounded-ms)
      (time-it iters (lambda () (run 1 (e) (str-bounded e depth) (str-matches e io-pairs)))))
    (printf "    bounded:    ~ams  ~a~n"
            (pad-right 8 (real->decimal-string bounded-ms 3))
            (fmt-result bounded-r))
    (flush-output)

    (define-values (bank-r bank-ms)
      (time-it iters (lambda () (run 1 (e) (str-bank e) (str-matches e io-pairs)))))
    (printf "    bank:       ~ams  ~a~n"
            (pad-right 8 (real->decimal-string bank-ms 3))
            (fmt-result bank-r))
    (flush-output)

    (define-values (bank-w-r bank-w-ms)
      (timed-run str-bank-w-budget-ms
        (lambda () (run-w 1 (e) (str-bank-w e) (str-matches e io-pairs)))))
    (printf "    bank-w:     ~ams  ~a~n"
            (pad-right 8 (real->decimal-string bank-w-ms 3))
            (fmt-result bank-w-r))
    (flush-output)

    (define-values (host-r host-ms)
      (time-it iters (lambda () (str-host-bank-result depth io-pairs))))
    (printf "    host-bank:  ~ams  ~a~n"
            (pad-right 8 (real->decimal-string host-ms 3))
            (fmt-result host-r))
    (printf "~n")
    (flush-output)))

(module+ main
  (arith-bench)
  (str-bench))
\end{minted}

\subsubsection{\texttt{bench/deep-bench.rkt}}
\label{sec:appendix-deep-bench}

\inputminted{racket}{supplementary/bench/deep-bench.rkt}

\subsubsection{\texttt{bench/order-bench.rkt}}
\label{sec:appendix-order-bench}

\inputminted{racket}{supplementary/bench/order-bench.rkt}

\subsubsection{\texttt{bench/decay-bench.rkt}}
\label{sec:appendix-decay-bench}

\inputminted{racket}{supplementary/bench/decay-bench.rkt}

\end{document}